# Measuring SES-related traits relating to technology usage: Two validated surveys

Chimdi Chikezie, Pannapat Chenpaiseng, Puja Agarwal, Sadia Afroz, Bhavika Madhwani, Rudrajit Choudhuri, Andrew Anderson, Prisha Velhal, Patricia Morreale, Christopher Bogart, Anita Sarma, Margaret Burnett

## ABSTRACT

Software producers are now recognizing the importance of improving their products' suitability for diverse populations, but little attention has been given to *measurements* to shed light on products' suitability to individuals below the median socioeconomic status (SES)—who, by definition, make up half the population. To enable software practitioners to attend to both lower- and higher-SES individuals, this paper provides two new surveys that together facilitate measuring how well a software product serves socioeconomically diverse populations. The first survey (SES-Subjective) is who-oriented: it measures who their potential or current users are in terms of their subjective SES (perceptions of their SES). The second survey (SES-Facets) is why-oriented: it collects individuals' values for an evidence-based set of facet values (individual traits) that (1) statistically differ by SES and (2) affect how an individual works and problem-solves with software products. Our empirical validations with deployments at University A and University B (464 and 522 responses, respectively) showed that both surveys are reliable. Further, our results statistically agree with both ground truth data on respondents' socioeconomic statuses and with predictions from foundational literature. Finally, we explain how the pair of surveys is uniquely *actionable* by software practitioners, such as in requirements gathering, debugging, quality assurance activities, maintenance activities, and fulfilling legal reporting requirements such as those being drafted by various governments for AI-powered software.

## 1. INTRODUCTION

Recently, software companies have begun casting their eyes on a relatively untapped market—people in the lower half of the socioeconomic status (SES) spectrum [Cusp Capital 2019]. For example, the Microsoft vision statement is now "to empower every person and every organization on the planet to achieve more" [Microsoft n.d.], half of whom are lower-SES people.

But how can software practitioners know whether their software products actually do empower a broad spectrum of SES-diverse users? The answer lies in measurement. Several kinds of measurement together are needed to answer this need, and this paper provides two of them, in the form of two validated survey instruments: an SES-Subjective Survey and an SES-Facets Survey.

The primary measurement device in this survey duo is the SES-Facets Survey, but for many of this survey's uses, it needs to be paired with the SES-Subjective Survey. And despite its supporting role, the SES-Subjective Survey makes contributions in its own right.

### 1.1 Measuring Subjective SES

The SES-Subjective Survey measures an individual's perception of their socioeconomic status as compared to others [Tan et al. 2020]. We developed it because in some cases—including our case of needing to validate the SES-Facets Survey—data on respondents' SES are needed along with data on their facets. Unfortunately, no existing SES-oriented survey was suitable for our goals, as we detail in Section 2.2, so we needed to create an SES-Subjective Survey of our own.

Because this survey measures (potential) users' SES distribution, software practitioners can use it to inform requirements gathering, for market research, for software product evaluations, and to produce any required SES-diversity reports. As one example, the survey might be used in conjunction with a user study to measure whether and how much participants' experiences with a software product or a software feature differ by their socioeconomic statuses.

The goal of the SES-Subjective Survey is not to measure an individual's SES according to some external standard. Rather, it aims to measure *subjective SES*—the SES with which an individual self-identifies internally—because of the strong ties between individuals' internal self-identities and their behaviors [Tan et al. 2020]. For example, individuals who see themselves as lower-SES tend toward lower-SES behaviors—which often persist even after increases to their (objective) SES by external thresholds such as income [Bian and Wu 2023, Browman et al. 2019, Destin 2019, Destin et al. 2017, Duan et al. 2022, Laurin and Engstrom 2020, Manstead 2018, Martin et al. 2016, Tan et al. 2020]. Because an individual's subjective perception of their own SES has such strong ties to their behaviors, we selected subjective SES[1] to facilitate measuring a software product's fit to SES-diverse users' technology-related behaviors.

In creating the SES-Subjective Survey we encountered several challenges. A particularly difficult challenge was devising a survey that could be used by software practitioners in most countries; however, many existing SES-oriented surveys rely on region/country-specific standards (e.g., ownership of objects that are luxuries in *that* region, etc.), rendering the survey useless to anyone outside of that region. Secondly, even if we had wanted to use income, many potential respondents are unwilling to provide their income information, to the detriment of the survey's response rate [Turrell 2000], which would render the survey's results potentially useless. A related challenge is that many companies will not even *ask* income information for privacy-related liability reasons, which would render the survey useless to companies unwilling to ask. We intend for our SES-Subjective Survey to avoid these challenges, which we frame as the following research question:

*RQ1-SES: What region/country/domain-agnostic indicators, which do not include income, reliably contribute to measuring the socioeconomic status (SES) to which an individual believes they belong?*

## 1.2 Beyond Subjective SES: Measuring SES Facets for Actionability

Whereas measurements of users' subjective SES in combination with user studies can point out problematic features that disproportionately affect users in different SES strata, that information alone is not fine-grained enough to provide actionability. For example, a user study combined with the SES-Subjective Survey may reveal that a feature displaying video viewing history was problematic to people in some SES levels—but what should developers *do* to improve this situation? To provide actionability, measurements need to reveal *why* these problems arose.

This "why" information is what the SES-Facets Survey provides. In our example, suppose the SES-Facets Survey reveals that users who find that feature problematic often have slow/unreliable internet access. To act upon this "why" information, a software practitioner might then decide to change the system to display history information only upon demand, or display it only when the system detects high internet bandwidth, or have history saving default to "off" with users needing to opt-in to save it, etc.

The SES-Facets Survey provides this why-oriented information by measuring SES-diverse individuals' traits (termed *facet values*, in this paper) that relate to individuals' behaviors with technology. Our SES-Facets Survey leverages Hu et al.'s literature analysis, which synthesized from extensive foundational literature a set of *facets* (traits) that (1) relate to problem-solving with technology and (2) have been shown to cluster by socioeconomic status [Hu et al. 2021].

An *SES facet* is a trait type with a range of possible values, and an *SES facet value* is the particular value an individual has for that facet. The SES facets are so named because their values cluster by individuals' SES. They are: *Technology Self-Efficacy; Attitudes toward Technology Risks; Perceived Control and Attitude Toward Authority;*

[1] In the remainder of this paper, we use the term "SES" to mean subjective SES, unless otherwise stated.

*Technology Privacy and Security; Communication: Literacy/Education/Culture;* and *Access to Reliable Technology* [Hu et al. 2021, Agarwal et al. 2023]. For example, *Access to Reliable Technology*'s facet values range from low access to high access to reliable technology, and these values statistically vary by SES. We describe the SES facets in more detail in Section 2.

Thus, we intend for our SES-Facet Survey to answer the following research question:

*RQ2-facets: What indicators reliably contribute to measuring individuals' SES facet values?*

## 1.3 Validations

From the outset, validation was part of our survey creation. The validations began with (1) multiple pilot sessions to refine the questions and to iteratively assess their suitability to SES-diverse participants.

We then (2) deployed the surveys at two universities with SES-diverse student populations. We analyzed the response data from participants using the SmartPLS v4 tool to find the indicator reliability and internal consistency reliability of the survey questions [SmartPLS 2024], thereby selecting an optimum set of questions to use. To also validate the SES-Subjective Survey's results against a known "ground truth", we then (3) validated the SES-Subjective Survey questions by comparing against the actual SES of participants at University A, which the university's Financial Aid office provided to us in a privacy-preserving form. These validation efforts answer the reliability aspects of RQ1 and RQ2.

Finally, (4) we validated our data against the foundational literature as another form of "ground truth". Specifically, we statistically tested whether the participants' SES identities related to their facet values in ways consistent with the literature, which we express as RQ3:

*RQ3-SESAndFacets: Did the distribution of our participants' facet values correspond with their subjective socioeconomic status (SES)?*

## 1.4 Contributions

Thus, the contributions of this paper are:

- An SES-Facets Survey (from RQ2), primarily to shed light upon the "why"s behind technology mismatches or potential mismatches to SES-diverse users.
  - *Uniqueness:* The SES-Facets Survey is the only survey to capture an individual's facet values that are both relevant to software usage and tied with SES.
  - *High resolution*: In so doing, SES-Facets produces high-resolution measurements of facets affecting how SES-diverse individuals go about problem-solving when using a software product's features.
  - *Actionability*: Because the SES-Facets Survey focuses on individuals' problem-solving processes and attributes, it can reveal the "why"s behind technology mismatches that disproportionately disadvantage people in a particular SES stratum. We present six use-cases that show how software practitioners can leverage this actionability.
- An SES-Subjective Survey (from RQ1), primarily for situations where SES statistics and/or SES reporting are needed. Although some software practitioners may find other existing SES surveys to be substitutable, our survey has unusual features particularly relevant to the software industry:
  - *Generality*: The SES-Subjective Survey is general—it is region/country-agnostic and domain-agnostic. Its generality enables use by software companies around the world, including those with global customer bases and those who offer products in multiple domains.
  - *Does not ask income*: The SES-Subjective Survey does not ask for income, which enables use by software companies whose policies are to avoid asking income information, and further avoids survey-response drops due to income-related questions.

- *Measures subjective SES*: The SES-Subjective Survey measures the SES to which the respondent *believes they belong*—not how they fit some external categorization scheme. People's self-identifications are, in general, closely tied with their behavior and this is also true of SES [Tan et al. 2020], which makes subjective SES particularly relevant to software practitioners interested in how their software products can provide a good fit to their users' technology *behaviors*.

- Validations: Both these surveys have been validated through pilots, multiple measurements of statistical reliability, and a validation that *our* participant response data used for these validations is indeed consistent with foundational literature.

# 2. BACKGROUND AND RELATED WORK

## 2.1 SES Facets

Our SES-Facets Survey measures the facets that Hu et al. developed and then Agarwal et al. refined. Hu et al. developed the facets by synthesizing over 200 research papers from multiple fields. From this, they derived a core set of five SES *facets*—types of individual traits, each with a range of possible values that statistically differ across different SES strata—that differently impact individuals' user experiences based on their SES strata [Hu et al. 2021]. Agarwal et al.'s field study then showed the facets to be effective at pinpointing inclusivity biases particularly impacting individuals in different SES strata, and also recommended splitting one of the facets into two, producing an updated total of six facets [Agarwal et al. 2023]. A brief summary of each of these six facets follows:

*Technology Self-Efficacy*: The Self-Efficacy facet refers to an individual's belief and confidence in their ability to effectively use technology. Studies show that, due to systemic disparities in access to technology, individuals from lower SES backgrounds are statistically more likely to have lower technology self-efficacy than individuals from higher SES backgrounds [Hatlevik et al. 2017; Vekiri 2010; Vekiri and Chronaki 2008]. Among self-efficacy's effects on an individual's ultimate success with a task are their attribution of the cause of difficulties they encounter (i.e., whether they believe they caused the difficulties or the system did), and their willingness to persevere in the face of difficulty and try different approaches to the problem if their first attempt fails [Bandera 1993]. Such effects have been reported multiple times with technology/computer self-efficacy (e.g., [Stumpf et al. 2020]).

*Attitudes toward Technology Risks*: The Risks facet captures an individual's tolerance levels or willingness to take risks when using technology. Lower-SES individuals, often challenged by the limitations of the technology available to them and/or limited time availability (e.g., juggling multiple jobs to make ends meet, relying on public transportation, etc.) are often cautious about using technologies/features that might not work out well for them, such as unfamiliar technologies [Antee 2021; Yardi and Bruckman 2012]. An individual's attitude toward such risks can affect the ways they use technology. For example, individuals with tolerance toward technology risks are more likely to try new features and experiment with different settings, even given the risk of the new features not being usable or useful. In contrast, risk-averse individuals may be more cautious when using technology and may be less likely to explore new features or try new apps. Note: In Hu et al.'s original work, this facet also included Privacy/Security risks, but Agarwal et al.'s field study showed the utility of splitting Privacy/Security off into a separate facet (shown later in this list), to encourage practitioners to consider risks beyond solely privacy and security risks [Hu et al. 2021, Agarwal et al. 2023].

*Perceived Control and Attitude Toward Authority*: The Control/Authority facet captures how much control different individuals perceive that they have over outcomes in their lives, which interrelates with how they respond to authority figures. Many lower-SES individuals perceive a lack of control over outcomes in their lives, including limited recourse to authority figures' decisions, and these attitudes can extend to technology as an "authority figure" and to its outcomes [Kalman 2014; Petro et al. 2020; Toorn et al. 2014]. An individual's perception of control over their outcomes and their attitude toward authority directly ties to their behaviors with and around technologies. For example, if individuals with perceptions of little recourse to authority figures' decisions regard computers as having authority over them, they can be more accepting of negative computer outcomes (e.g., web pages with greyed out "submit" buttons, paths leading to error messages, etc.) [Lee and Ko 2011] than people who are more critical/questioning of authority.

*Technology Privacy and Security*: The Privacy and Security facet captures an individual's concerns about their technological privacy and security. Given the higher prevalence of police and government surveillance in lower-SES communities [APA 2017] and the high usage of shared/public devices among lower-SES individuals [Sambasivan et al. 2018], lower-SES individuals tend to exhibit heightened caution regarding privacy and security. For example, Black Americans, who are disproportionately lower-SES, are more likely than White Americans to worry that the government is tracking them online [Auxier et al. 2019]. These differences can affect technology usage in numerous ways. For example, some lower-SES individuals avoid technology risks by avoiding engaging with technology whenever possible, and others control the amount of information available online by deleting it regularly and extensively [Brayne 2014; Sambasivan et al. 2018].

*Communication: Literacy/Education/Culture*: The Communication facet captures an individual's comfort level in communicating with technology in whatever language/vocabulary/communications devices the technology dictates—whether it be college-level English, Spanish, techie jargon, custom icons, etc. The facet's full name, "Communication: Literacy, education, and cultural background", serves to emphasize the myriad factors that can affect someone's communication ability with that technology's communications. Socioeconomic differences can arise in this facet due to inequities in educational opportunities, diverse cultural backgrounds, disparities in literacy levels, etc., and any of these differences can significantly influence users' success in using the technology's features [Aesaert and Braak 2015; Cheung 2017; Guo 2018].

*Access to Reliable Technology*: This facet encompasses both an individual's overall access to technology and the reliability and availability of that technology. Examples include reliable internet, access to transportation to travel to a shared technology location (e.g., to use the library's computers), up-to-date hardware, up-to-date software, etc. Access to reliable devices with reliable internet connection can be impacted by device ownership, sharing, and reliance on public devices, all of which vary across SES strata and can influence individuals' approaches to using technology, their learning processes, and their overall user experiences [Guberek et al. 2018; Margolis et al. 2017; Ryan 2018]. For example, individuals with low access to reliable technology may not be able to take advantage of some technology features.

Although extensive foundational research exists on these facets (see [Hu et al. 2021] for a summary), no validated survey exists that includes these six facets.

## 2.2 SES-Related Surveys

As mentioned earlier, to validate our SES-Facets Survey we needed to collect respondents' subjective SES. Although we had expected to be able to reuse an existing SES-related survey, an extensive search of the literature did not reveal any existing SES-related survey that satisfied our criteria. The criteria were (criterion 1): the survey questions should not ask respondents to specify their income, because many people are unwilling to answer income questions and some organizations are unwilling to collect people's income data; (criterion 2): the survey questions should not be country/region-specific or domain-specific; and (criterion 3): the survey should measure an individual's subjective perception of their SES, not their SES according to external categorization (e.g., using income). In the end, despite finding numerous works about SES-related surveys (e.g., [Pradhan et al. 2018; Roohafza et al. 2021; Shafiei et al. 2019], and the 41 others we describe in this section and in Section 3.1), none fulfilled these criteria.

A detailed example of one type of survey that did not fulfill these criteria can be seen in Easterbrook et al.'s analysis of SES impacts on an individual's access to resources, social status, and personal opportunities that shape self-concepts [Easterbrook et al. 2019]. They analyzed data from two large UK surveys—the UK Household Longitudinal Study [Univ. Essex 2018] and the UK's Citizenship Survey [UK Dept. Communities and Local Government 2012]. The data from these surveys consisted of participants' responses to household/personal income levels, along with other SES indicators like social class. Participants were categorized into low and high SES based on a composite measure of annual income, education, and occupation. Since these surveys collect income information, they do not fulfill our criterion 1. Another issue, relating to criterion 3, is that the surveys used in this study measure the participants' SES according to some external standard, whereas our survey requirement was to capture participants' *own* assessments of their SES.

Even after removing surveys such as those described above and others that directly asked respondents for their income (criterion 1), e.g., [Adamu et al. 2018; Le et al. 2015; Pradhan et al. 2018; Raihan et al. 2017; Shafiei et al. 2019], many remaining SES-related surveys that we located were domain-focused, thus not fulfilling criterion 2. For example, Patel et al. surveyed ownership of certain household assets as SES indicators relevant to health outcomes of rural women and children [Patel et al. 2020]. Goodwin et al. discussed SES indicators that can affect common mental disorders related to unemployment and low education [Goodwin et al. 2018]. Raihan et al. included different ways to identify food security status to focus on malnutrition of children [Raihan et al. 2017]. Jordan et al. emphasized maternal education as an indicator impacting incidence and outcomes of childhood arterial ischemic stroke [Jordan et al. 2018]. Carrasco-Escobar et al. surveyed wealth and maternal education associated with preventive measures for malaria infection [Carrasco-Escobar et al. 2021]. In the education domain, surveys such as Yang et al.'s inquired about possession of personal items such as story books and other household items as indicators of SES to find relationships with students' reading achievement [Yang and Gustafsson 2004]. Mustofa et al. used prestige, power, and authority questions for measuring SES to find relationships with types of publications among graduate students [Mustofa et al. 2019]. Outside these domains, other examples of SES-related surveys that were domain-focused include surveys relating to lexical comprehension of toddlers [Rosemberg and Alam 2021], marketing research [Kasemsant et al. 2019], neighborhood design [Suwannasang 2019], and food security [Babatunde et al. 2007; Yadegari et al. 2017].

Besides domain specificness, several surveys did not fulfill our criterion 2 because they included questions that turned out to be region- or country-specific. Examples include asking about ownership of household/personal items (e.g., tv, refrigerator, air conditioner, personal computer, car, etc.), water/sanitation, residential location, household size, personal education, individual's occupation, etc. [Omer and Al-Hadithi 2017]. To illustrate, consider a survey question about owning a refrigerator (one of the questions we found). Although this question could differentiate lower-SES from higher-SES individuals in some countries (e.g., fewer than half of household's own refrigerators in Afghanistan, Angola, Bangladesh, Bolivia, Cambodia, Ethiopia, and other countries [Institute for Management Research n.d.]), it would not do so in many others. For example, as of 2018, over 99% of American households owned a refrigerator [Hammond 2018], so questions about refrigerators would have little value in measuring someone's SES in such countries.

Conceptually close to our SES-Subjective Survey is the MacArthur Scale of Subjective Social Status (SSS). Adler et al. pioneered the construct of Subjective Social Status, an individual's perception of their social status [Adler et al. 1994, Adler et al. 2000]. The MacArthur Scale of SSS measures participants' own perception of where they believe they "stand" in terms of being valued in their country or community [Stanford SPARQtools n.d.]. It is a single-question survey: it includes a picture of a ladder (Fig. 1), and gives the following instructions (elided for brevity):

> *"Think of this ladder as representing where people stand in <country/community>. At the top of the ladder are the people who are best off—those who have the most money, most education, and most respected jobs. At the bottom...Where would you place yourself on this ladder? Please place a large 'X' on the rung..."* [Stanford SPARQtools n.d.]

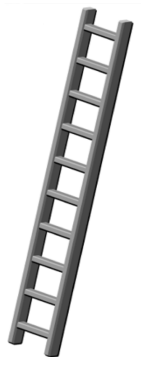

**Fig. 1** The ladder used in the MacArthur Scale of Subjective Social Status [Adler et al. 2000]

MacArthur Scale one-question surveys have been used widely for both adults and adolescents, particularly within the U.S. For example, they have been used to study diverse populations within the U.S., such as with African

Americans, Asian-Americans, Chinese-Americans, European-Americans, Latinx-Americans, Mexican-Americans, Pacific Islanders, and various immigrant populations [Bullock and Limbert 2003, Franzini & Fernandez-Esquer 2006, Garza et al. 2017, Goodman et al. 2001, Leu et al. 2008, Ostrove et al. 2000, Stanford SPARQtools n.d.]. Although the MacArthur Scale of SSS has been shown to be robustly associated with health and well-being, and some use it to measure subjective SES [Tan et al. 2020], it has been shown to not be a measure of subjective SES alone [Galvan et al. 2023]. As Galvan et al. showed, it confounds two distinct concepts: subjective SES and perceived standing in the community or society, with the latter concept being the most strongly associated with the survey scoring [Galvan et al. 2023]. Thus, it does not satisfy criterion 3.

Since none of the above surveys or any others that we could locate were able to fulfill all three of our criteria, we developed our own SES-Subjective Survey, using the methods we detail in the next section.

# 3. The SES-Subjective Survey

Beyond its role in validating the SES-*Facet* Survey, the SES-Subjective Survey has particular strengths that may make it especially well-suited to the software industry.

The goal of the SES-Subjective Survey is to measure the socioeconomic status to which an individual *subjectively believes* they belong, without directly asking for their income, in ways that are both domain-agnostic and region/country-agnostic. The development of this survey was guided by the following research question:

*RQ1-SES: What region/country/domain-agnostic indicators, which do not include income, reliably contribute to measuring the socioeconomic status (SES) to which an individual believes they belong?*

We designed this survey by collecting numerous objective SES and subjective SES questions from the literature, which we then refined iteratively in combination with pilots, and then deployed the survey at University A, as we detail in Section 3.1. We then validated the SES-Subjective Survey by both evaluating the reliability of the questions and validating it with the "ground truth" of respondents' known objective SES, as we detail in Section 3.2.

## 3.1 Design and Deployment of the SES-Subjective Survey

We began by collecting indicators that foundational research has found to matter for SES. Because our aim was completeness (not counting things, as in systematic literature studies), we did not constrain our search in any way. For example, there were no restrictions to particular search strings, particular databases, particular fields of study, particular date ranges, etc. Instead, we searched widely across multiple fields, and when the paper was particularly useful, we also followed forward and backward citations. For example, with Psaki et al.'s cross-sectional study of eight resource-limited settings in multiple countries [Psaki et al. 2014], we followed forward and backward citations to that study. We harvested the indicators and, when available, survey questions for these indicators. We grouped these indicators (and their questions) into 13 categories and whether the literature considers the indicator(s) to be more a causation or more a reflection (symptom) of SES. For example, the Education category is widely regarded a cause of SES, so we marked it as "Causation" in Table 1, and included in it the following SES indicators we found: parent's education, self-education, couple's education, household education. Table 1 shows all the resulting indicator categories sorted on the type of the indicator (causation or reflection) and the number of questions from each of these indicators that we eventually selected (discussed later). Appendix B provides full details for each category.

From this list, we selected categories to include in our survey based on the following criteria: (1) each indicator in a category was supported by at least two research papers, (2) SES indicators and their questions did not directly ask income, and (3) they were not culture or region/country or domain specific. For example, we dropped questions related to access to healthcare [Reibling et al. 2019] or transportation [Kaiser and Barstow 2022], as these are public services in some countries and readily available, but not so in other countries. Similarly, household size can be culture dependent [Esteve and Liu 2017]; some cultures prefer nuclear families, while others live in multi-generational families. This filtering resulted in the 7 categories shown in the top section of Table 1.

Table 1: Categories for the SES-Subjective Survey questionnaire development (see Appendix B for details)

| SES indicator type | Category | # of selected questions | Paper references |
|---|---|---|---|
| SES indicator categories we selected for the survey | | | |
| Causation | Demographics | 1 | [Babatunde et al. 2007, Omer and Al-Hadithi 2017, Porter 2016, Shafiei et al. 2019, Suwannasang 2019, Yadegari et al. 2017] |
| Causation | Education | 2 | [Adamu et al. 2018, Ali and Bakar 2019, Babatunde et al. 2007, Carrasco-Escobar et al. 2021, Goodwin et al. 2018, Jordan et al. 2018, Lee and Yi 2021, Mustofa et al. 2019, Omer and Al-Hadithi 2017, Porter 2016, Pradhan et al. 2018, Raihan et al. 2017, Roohafza et al. 2021, Rosemberg and Alam 2021, Shafiei et al. 2019, Suwannasang 2019, Yadegari et al. 2017] |
| Causation | Employment | 1 | [Goodwin et al. 2018, Lee and Yi 2021, Omer and Al-Hadithi 2017, Porter 2016, Roohafza et al. 2021, Shafiei et al. 2019, Yadegari et al. 2017] |
| Reflection | Food Security | 1 | [Raihan et al. 2017, Yadegari et al. 2017] |
| Reflection | Housing Situation | 3 | [Adamu et al. 2018, Colston et al. 2020, Goodwin et al. 2018, Jordan et al. 2018, Kasemsant et al. 2019, Le et al. 2015, Lee and Yi 2021, Patel et al. 2020, Porter 2016, Roohafza et al. 2021, Rosemberg and Alam 2021, Shafiei et al. 2019, Yadegari et al. 2017] |
| Reflection | Standard of Living | 2 | [Adamu et al. 2018, Babatunde et al. 2007, Kasemsant et al. 2019, Lee and Yi 2021, Omer and Al-Hadithi 2017, Patel et al. 2020, Porter 2016, Pradhan et al. 2018, Raihan et al. 2017, Roohafza et al. 2021, Shafiei et al. 2019, Yadegari et al. 2017, Yang and Gustafsson 2004] |
| Reflection | Water/Sanitation | 1 | [Colston et al. 2020, Porter 2016. Pradhan et al. 2018] |
| SES indicator categories **not** selected for the survey | | | |
| Causation | Household Size | 0 | [Babatunde et al. 2007, Coleman and Fulford 2022, Colston et al. 2020, Suwannasang 2019, Yadegari et al. 2017] |
| Causation | Income | 0 | [Adamu et al. 2018, Coleman and Fulford 2022, Goodwin et al. 2018, Jordan et al. 2018, Le et al. 2015, Lee and Yi 2021, Mustofa et al. 2019, Pradhan et al. 2018, Raihan et al. 2017, Shafiei et al. 2019, Suwannasang 2019, Yadegari et al. 2017] |
| Causation | Occupation | 0 | [Ali and Bakar 2019, Goodwin et al. 2018, Mustofa et al. 2019, Omer and Al-Hadithi 2017, Porter 2016, Rosemberg and Alam 2021, Suwannasang 2019] |
| Reflection | Healthcare | 0 | [Adamu et al. 2018, Lee and Yi 2021] |
| Reflection | Transportation | 0 | [Adamu et al. 2018] |
| Reflection | Well-being | 0 | [Adamu et al. 2018] |

For each category, we evaluated the SES indicator and the questions for measuring these indicators when available. To keep the survey size viable, we included at least one SES indicator per category and removed those that were not well-supported as SES indicators by foundational literature. For example, although many research works have found a connection between parents' education level and SES, the connection between a partner's education to SES is not as well connected [Ali and Bakar 2019; Carrasco-Escobar et al. 2021; Colston et al. 2020; Jordan et al. 2018; Pradhan et al. 2018; Raihan et al. 2017; Rosemberg and Alam 2021]. Thus, we kept personal and parent's education but dropped indicators (and their questions) for couples/household education.

We then analyzed the questions for the selected indicators and removed questions that were duplicates covering approximately the same SES indicator. For example, the questions "I have enough decent food to eat" and "My food quality is insufficient compared to others in my <state/country>" convey similar ideas and keeping both was considered (approximately) redundant; hence, we kept the former only. Next, we removed questions that were region-specific or domain-specific. Examples of region-specific questions have been given above; an example of domain-specific questions is ownership of specific personal items relating to a particular domain (e.g., owning story books as an SES indicator relating to reading achievement of students [Yang and Gustafsson 2004]).

We then piloted the remaining questions multiple times to remove questions that pilot participants said did not make sense to them. The final set of 11 SES-Subjective Survey questions (S1, S2…) spanning the 7 categories are listed in Table 2.

Table 2: The 11 questions in the SES-Subjective Survey as deployed at University A.

| SES-Subjective Survey Questions | Scales | Category |
|---|---|---|
| S1. I have a higher education level compared to everyone else in my permanent state/country. | 1 (Completely Disagree) to 9 (Completely Agree) | Education |
| S2. My parents have a higher education level compared to everyone else in my permanent state/country. | 1 (Completely Disagree) to 9 (Completely Agree) | Education |
| S3. I'm a minority in my permanent state/country. | 1 (Completely Disagree) to 9 (Completely Agree) | Demographics |
| S4. My housing is crowded compared to everyone else in my permanent state/country. | 1 (Completely Disagree) to 9 (Completely Agree) | Housing Situation |
| S5. My neighborhood is safe | 1 (Completely Disagree) to 9 (Completely Agree) | Housing Situation |
| S6. What is your current housing status now? | 1. I own my residence (or my family does, and I live there).<br>2. I live in rented housing.<br>3. I don't have stable housing. | Housing Situation |
| S7. What is your current job status now? | 1. Employed full-time (either self-employed or by an employer)<br>2. Employed by multiple part time jobs (either self-employed or by an employer)<br>3. Employed part-time (either self-employed or by an employer)<br>4. Otherwise supported (no need for employment)<br>5. Unemployed | Employment |
| S8. I have the ___ amount of wealth of everyone in my permanent state/country | 1 (Least Wealthy) to 9 (Most Wealthy) | Standard of Living |
| S9. My household makes enough to live comfortably. | 1 (Never) to 9 (Always) | Standard of Living |
| S10. I have enough decent food to eat. | 1 (Never) to 9 (Always) | Food Security |
| S11. I have access to clean water for my needs. | 1 (Never) to 9 (Always) | Water Sanitation |

**Deployment at University A.** We deployed the SES-Subjective Survey as per Table 2 to students at University A, with the help of the Office of Financial Aid. Because the Office of Financial Aid handled the deployment, we knew neither the identities nor the identifying email addresses of the participants. The Office of Financial Aid distributed the survey to students according to the following criteria:

*Lower-SES*: To be in this group, students had to be undergraduates who (1) had previously provided income/support information to the university, (2) showing that they were eligible for a Pell Grant, a federal financial aid program for students who are below the U.S. poverty line (7,309 students at that time).

*Higher-SES*: To be in this group, students had to be undergraduates who (1) had previously provided income/support information to the university, (2) showing that they were not eligible for *any* need-based scholarships (4,787 students at that time).

During this deployment, we combined the SES-Subjective Survey questions with the SES-Facets Survey's questions to pilot the latter, the design of which is discussed in Section 4.1. We counterbalanced the order in which the SES-Subjective (S) and SES Facet (F) questions appeared (Table 3) in the surveys the Office of Financial Aid sent out.

Table 3: Counterbalanced order of SES-Facets survey questions and SES-Subjective survey questions deployed to Higher-SES and Lower-SES participants. The entries in the cells are the way we refer to each version of the combined surveys.

| | S questions first, then F questions | F questions first, then S questions |
|---|---|---|
| Higher SES (H) | H-SF | H-FS |
| Lower SES (L) | L-SF | L-FS |

The survey was sent to 12,096 students in total and we compensated the respondents via a raffle of a $50 Amazon gift card for every 50 respondents who opted into being part of the raffle.

A total of 464 students responded to the surveys, 284 lower-SES students (L-FS and L-SF) and 180 higher-SES students (H-FS, H-SF). The response rate was 3.84% which is below Smith et al.'s recommended response rate of >=6% [Smith et al. 2013]. We return to this point in Section 5.2.

## 3.2 Validation of the SES-Subjective Survey

To validate the SES-Subjective Survey, we first evaluated the reliability of the survey questions in measuring subjective SES, and then validated the survey by comparing the survey responses with a "ground truth" of respondents' known objective socioeconomic statuses.

### *3.2.1 Reliability Measurements*

We used the SmartPLS v4 tool [Sarstedt and Cheah 2019] to test for indicator reliability and internal consistency reliability. Indicator reliability seeks to assess the relationship between indicators (questions in the survey) and their corresponding constructs [Kock 2014]. Constructs (e.g., subjective SES) are unobservable measures that are captured using one or more observable indicators (e.g., food security, housing, employment).

We assessed indicator reliability using the indicators' factor loadings, measuring the extent to which an individual indicator (question) measures its corresponding construct (subjective SES). Factor loadings of 0.6 and higher are considered adequate [Hair et al. 2019] for exploratory studies, so we used this as our threshold.

We calculated the factor loadings of the 11 SES-Subjective Survey questions and discarded S1, S3-S7, and S11, since they fell below the threshold. This left us with four questions: S2, S8, S9, and S10. See Table 4 for the factor loadings for these questions. Three of them (S8, S9, and S10) demonstrated indicator reliability above the 0.6 threshold [Hair et al. 2019, Russo and Stol 2021], indicating that these questions reliably measure the subjective SES construct.

Table 4: Indicator Reliability and Internal Consistency Reliability Result of SES-Subjective Survey. (S2 was kept because of theory.)

| SES-Subjective Question | Composite Reliability | Factor Loading | Outer Weights |
|---|---|---|---|
| S8: I have the ___ amount of wealth of everyone in my permanent state/country. | 0.81 | 0.79 | 0.39 |
| S9: My household makes enough to live comfortably. | | 0.89 | 0.40 |
| S10: I have enough decent food to eat. | | 0.82 | 0.39 |
| S2: My parents have a higher education level compared to everyone else in my permanent state/country. | | 0.30 | 0.06 |

Despite S2's low factor loadings, we retained it because literature has found "parent's education" to be an important factor impacting SES [Ali and Bakar 2019; Carrasco-Escobar et al. 2021; Colston et al. 2020; Jordan et al. 2018; Pradhan et al. 2018; Raihan et al. 2017; Rosemberg and Alam 2021]. This is in line with prior research [Friedenberg 1995; Hair 2019] stating that, regardless of statistical eligibility, items can be retained based on strong theoretical relevance and contribution to the underlying construct.

We then used the Composite Reliability test to measure internal consistency reliability. Internal consistency reliability is the extent to which the different indicators (questions) are consistent with one another and can reliably and consistently measure the same construct (subjective SES). An acceptable level of reliability indicates that the indicators (questions) are referring to the same construct. Values in the range of 0.6-0.95 are within the accepted range for composite reliability. The SES-Subjective questions achieved a composite reliability score of 0.81 (Table 4), indicating that collectively, the four SES-Subjective questions (including S2), reliably and consistently measure the subjective SES construct.

### *3.2.2 Validation of the SES-Subjective Survey as a Proxy for SES*

A final validity check is to consider the reliability of the SES-Subjective Survey responses as a reasonable proxy for SES, by which we mean a reliable differentiator of the lower-SES participants from the higher-SES participants in our dataset. To do this check, we compared the university-provided ground truth (representing *objective* SES, as per the university's categorization described in Section 3.1) with our survey responses, which reflect participants *subjective* SES. If our survey results are valid, they should show a good relationship between the university-calculated objective SES and our survey's subjective SES, as in other research into objective vs. subjective SES [Tan et al. 2020]. Thus, we compared participants' survey responses with their known, objective SES as follows.

First, we computed each participant's S-score using weighted sums of each participants' responses to the four SES-Subjective questions:

$$S - score = w_1 * S2 + w_2 * S8 + w_3 * S9 + w_4 * S10$$

Each $w_i$ is the outer weight for that question from Table 4, which indicates each question's contribution to measuring SES. We used a two-sample t-test to compare the objectively lower-SES participants' S-scores with the higher-SES participants' S-scores. The results were that the objectively higher-SES participants had higher S-scores than the objectively lower-SES participants; these differences were statistically significant ($t=11$, $df=438.28$, $p = 2.2e-16$, two tailed t-test) with a large[1] effect size (Cohen's $d = 0.996$).

From this we conclude that the final SES-Subjective Survey in Table 4 is an effective differentiator of higher-SES participants from lower-SES participants, and a reliable proxy for the subjective SES construct.

# 4. The SES-Facets Survey

## 4.1 Design and Deployment

*RQ2-facets: What indicators reliably contribute to measuring individuals' SES facet values?*

The design of the SES-Facets Survey followed a similar approach as with the SES-Subjective Survey. We began by searching existing surveys and questionnaires (with a preference for validated surveys) that measured the concepts closely related to any of Section 2.1's six SES facets.

Two of these six facets, Self-Efficacy and Risk, are also used by the GenderMag method [Burnett et al. 2016]. GenderMag is an inspection method used to help software teams design gender-inclusive software by addressing five problem-solving facets that cluster by gender. Our questions on Self-Efficacy and on Risk originated from their respective facet questions in the validated GenderMag survey [Hamid et al. 2024], although our validation process resulted in refinements and reductions of the questions to a subset.

[1] We considered Cohen's d=0.2-0.5 to be a small effect, d=0.5-0.8 to be a medium effect, and d >0.8 to be a large effect as per convention [Cohen 1988].

We informed the questions for the rest of the facets by drawing from a wide literature base, but found the following papers particularly helpful: Bobak et al.'s questionnaire for the Perceived Control and Attitude Toward Authority facet [Bobak et al. 1998]; Chignell et al.'s and Parsons et al.'s validated questionnaires for the Technology Privacy and Security facet [Chignell et al. 2003, Parsons et al. 2017], Pishghadam et al.'s questionnaire for the Communication: Literacy/Education/Culture facet [Pishghadam and Zabihi 2011], and Renuka and Gurunathan and Rahim et al. 's questionnaires for the Access to Reliable Technology facet [Renuka and Gurunathan 2017; Rahim et al. 2020].

This process produced 147 facet-relevant questions. From this pool, we removed domain and country-specific questions, redundant questions, and questions not likely to vary by SES, while including questions for each facet's theoretical underpinnings [Agarwal et al. 2023, Hu et al. 2021].

We then refined the questions through several pilot studies. We first piloted the survey with a few community college and university students in Oregon and Tennessee. The second pilot was with attendees at a meeting for agriculture workers in Washington state; the third with residents of a "food desert" neighborhood in Indiana; and the fourth was with a few students in Oregon, to help us further clarify wording. The final pilot was with undergraduate students at University A (as discussed in Section 3.1). These pilots had sizable participation by people from lower-SES backgrounds.

After every pilot survey, we used participants' feedback to improve the questionnaire clarity. We also went back to the literature to ensure that we did not change the question wording too much. For example, the "Communication: Literacy/Education/Culture" facet includes three "subfacets", but our clarifying and editing had reduced it to mainly literacy. Thus, we re-added enough questions from the literature to cover all three. We piloted refined questions iteratively, ultimately producing the 32 questions listed in Appendix A.

We then deployed Appendix A's facet questions, along with Table 4's four-question SES-Subjective Survey, at University B, a university with wide cultural diversity. At the time of deployment, University B had 17,000 students enrolled from over 82 countries. As before, the two surveys were counterbalanced, with half the participants getting the SES-Subjective Survey first and then the SES-Facets Survey, and the rest of the participants getting the surveys in the opposite order.

The survey was open for six weeks. As before, we compensated participants via a raffle of a $50 Amazon gift card for every 50 respondents. A total of 565 students responded, a response rate of 3.32%. (Section 5.2 further discusses the response rate.) We then did a response validation check and deleted 39 duplicate[1] responses. We also deleted four other participants, three of whom had answered every question with the same Likert response option, (including questions that contradicted each other), and the fourth of whom did so on every question except one. This left 522 responses.

## 4.2 Validation of the SES-Facets Survey

We validated the SES-Facets Survey, using the same statistical procedures to calculate indicator reliability and internal consistency reliability as with the SES-Subjective Survey, using the 6 facets as 6 constructs instead of the subjective SES construct. Using these procedures, we started with all the questions from the deployed SES-Facets Survey of Appendix A, and then discarded the questions that did not meet the 0.6 factor loading threshold, with three exceptions for theoretical reasons. In this section, we discuss only those questions that remained.

### *4.2.1 Indicator Reliability and Internal Consistency Reliability*

We begin with the first construct, **Technology Self-efficacy**. From the statistical validation, this facet resulted in the 4 questions in Table 5. As the table shows, the factor loading for every question was above the 0.6 threshold, which means each question adequately reflects the Technology Self-efficacy facet construct. Further, these 4 questions together had a composite reliability of 0.92, which is within the acceptable range for internal consistency. Therefore, these four questions are reliable and consistent measures of Technology Self-efficacy.

[1] Participants could optionally provide email addresses they wanted us to use in the optional raffle. We detected the 39 duplicates by comparing those email addresses.

Table 5: Result of Technology Self-efficacy (composite reliability, factor loading, outer weight: see Section 3.2.1).

| Facet: Composite Reliability | Question | Factor Loading | Outer Weight |
|---|---|---|---|
| Technology Self-efficacy: 0.92 | F1a: I am able to use specialized software when I have just the built-in help for assistance. | 0.75 | 0.31 |
| | F1b: I am able to use specialized software when I have seen someone else using it before trying it myself. | 0.88 | 0.30 |
| | F1c: I am able to use specialized software when someone else has helped me get started. | 0.90 | 0.30 |
| | F1d: I am able to use specialized software when someone shows me how to do it first. | 0.88 | 0.28 |

The **Attitudes toward Technology Risks** facet statistical validation produced adequate factor loadings for F2 and F3, but not for F4 (Table 6). Despite this, we kept F4 in the survey for theory reasons: this aspect of risk is well-grounded in SES foundational literature [Hu et al. 2021]. These three questions together gave a composite reliability of 0.73, showing a satisfactory level of internal consistency reliability. (Recall from Section 3.2.1, values in the range of 0.6-0.95 are the allowable ranges for composite reliability [Hair et al. 2019]).

Table 6: Result of Attitudes toward Technology Risks (composite reliability, factor loading, outer weight: see Section 3.2.1).

| Facet: Composite Reliability | Question | Factor Loading | Outer Weight |
|---|---|---|---|
| Attitudes toward Technology Risks: 0.73 | F2: I avoid "advanced" features or "options" in specialized software. | 0.68 | 0.40 |
| | F3: I avoid activities when using specialized software that are dangerous or risky. | 0.87 | 0.64 |
| | F4: Despite the risks, I use features in specialized software that haven't been proven to work. | 0.49 | 0.35 |

The **Perceived Control and Authority** facet comprised two sub-constructs: Perceived Control (F5, F8) and Attitude toward Authority (F7, F10). As Table 7 shows, the factor loadings for the four questions are adequate, and both sub-constructs separately demonstrate a level of internal consistency within the allowable range.

Table 7: Result of Perceived Control and Authority (composite reliability, factor loading, outer weight: see Section 3.2.1). Each question showed reliability (factor loading) and for both sub-constructs, the questions showed internal consistency (composite reliability)

| Facet | Composite Reliabilities | Question | Factor Loading | Outer Weight |
|---|---|---|---|---|
| Perceived Control and Authority | Perceived Control: 0.81 | F5: I feel I have control over what happens in most situations. | 0.83 | 0.60 |
| | | F8: I feel I have control over what happens when I work with technology. | 0.83 | 0.61 |
| | Attitude toward Authority: 0.79 | F7: I often have the feeling that I am being treated unfairly. | 0.95 | 0.83 |
| | | F10: I often have the feeling that technology ends up treating me unfairly. | 0.64 | 0.34 |

As Table 8 shows**,** the **Technology Privacy, and Security** facet's questions F12 and F14 demonstrated adequate factor loadings (>0.6). We also included F11 for strong theory reasons relating to lower-SES individuals' previous negative experiences from being under surveillance [Benjamin 2019; Brown 2015; Lu et al. 2023]. These three questions had a composite reliability of 0.71, which shows internal consistency reliability.

F13's factor loading (0.35) was below the allowable threshold. We made it optional, because some researchers might choose to add it in their survey for theoretical reasons and some might not. It does have theory backing, but not as much as F11. For example, Marwick et al. found in their study that the practice of deleting pictures in social media sites was widespread among young people of lower-SES [Marwick et al. 2017]. Similarly, Sambasivan et al. found that people who shared devices managed their privacy by deleting their conversation and query information from the device [Sambasivan et al. 2018]. Including F13 in the survey gives a composite reliability of 0.69 which is within the allowable range.

Table 8: Result of Technology Privacy, and Security (composite reliability, factor loading, outer weight: see Section 3.2.1).

| Facet | Composite Reliabilities | Question | Factor Loading | Outer Weight |
|---|---|---|---|---|
| Technology Privacy, and Security | Considering F11, F12, F14: 0.71 | F11: Video cameras in public places that police can view help me feel safe in isolated places/night. | 0.51 | 0.41 |
| | | F12: I do not mind using my full name on social media. | 0.80 | 0.60 |
| | | F14: I do not want pictures of me on the internet or social media. | 0.67 | 0.44 |
| | Adding F13 to the other three: 0.69 | F13: I delete messages/posts/data from my mobile/social media to guard my personal data. (Optional) | 0.35 | 0.15 |

The **Communication: Literacy/Education/Culture** facet had four questions which we divided into two sub-constructs, Communication and Education. As shown in Table 9, the factor loadings of the questions in the groups were adequate. Further, the two sub-constructs' questions each had internal consistency reliability within the allowable range.

Table 9: Result of Communication: Literacy/Education/Culture (composite reliability, factor loading, outer weight: see Section 3.2.1).

| Facet | Composite Reliabilities | Question | Factor Loading | Outer Weight |
|---|---|---|---|---|
| Communication: Literacy/ Education/ Culture | Communication: 0.79 | F15: I usually understand materials referring to USA culture, politics, celebrities, and traditions, (e.g., "it crashed", "ROI", "cutting corners", "FYI", "driving me up the wall"...). | 0.85 | 0.68 |
| | | F20: I can speak English well. | 0.76 | 0.56 |
| | Education: 0.83 | F18: When growing up, my school provided me with high quality education. | 0.88 | 0.65 |
| | | F19: When growing up, my school gave me lots of technology to use. | 0.81 | 0.53 |

The **Access to Reliable Technology** facet had four questions which we divided into two parts, Technology Sharing and Reliable Technology. As shown in Table 10, the factor loadings of all the questions in the groups were adequate and both the subgroups were within the allowable internal consistency reliability range.

Table 10: Result of Access to Reliable Technology (composite reliability, factor loading, outer weight: see Section 3.2.1).

| Facet | Composite Reliabilities | Question | Factor Loading | Outer Weight |
|---|---|---|---|---|
| Access to Reliable Technology | Technology Sharing: 0.72 | F22: I use a SHARED mobile device or computer (e.g. shared with family, at public library, etc.) | 0.61 | 0.49 |
| | | F25: I have (for MY USE ALONE) this many mobile devices or computers in my household: (dropdown 0/1/2/more than 2)") | 0.87 | 0.80 |
| | Reliable Technology: 0.89 | F23: The mobile devices or computers that I usually use are reliable") | 0.88 | 0.50 |
| | | F24: I can get access to internet connection(s) that are reliable enough for my purposes. | 0.92 | 0.61 |

The final validated SES-Facets Survey questions set is shown in Table 11.

Table 11: Final list of Reliable SES-Facets Survey Questions after Analysis. For all questions except F25, response scales are from 1 (Completely Disagree) to 9 (Completely Agree). For F25, dropdown options are added with the question.

| Facet | Questions |
|---|---|
| Technology Self-efficacy | F1. I am able to use specialized software when...<br>a. I have just the built-in help for assistance.<br>b. I have seen someone else using it before trying it myself.<br>c. someone else has helped me get started.<br>d. someone shows me how to do it first. |
| Attitudes toward Technology Risks | F2. I avoid "advanced" features or "options" in specialized software.<br>F3. I avoid activities when using specialized software that are dangerous or risky.<br>F4. Despite the risks, I use features in specialized software that haven't been proven to work. |
| Perceived Control and Authority | F5. I feel I have control over what happens in most situations.<br>F7. I often have the feeling that I am being treated unfairly.<br>F8. I feel I have control over what happens when I work with technology.<br>F10. I often have the feeling that technology ends up treating me unfairly. |
| Technology, Privacy, and Security | F11. Video cameras in public places that police can view help me feel safe in isolated places/night.<br>F12. I do not mind using my full name on social media.<br>F14. I do not want pictures of me on the internet or social media.<br>F13. I delete messages/posts/data from my mobile/social media to guard my personal data. (Optional) |
| Communication: Literacy/ Education/ Culture | F15. I usually understand materials referring to <country's> culture, politics, celebrities, and traditions, (e.g., <example country's idioms, cultural references, ...>).<br>F18. When growing up, my school provided me with high quality education.<br>F19. When growing up, my school gave me lots of technology to use.<br>F20. I can speak English well. |
| Access to Reliable Technology | F22. I use a SHARED mobile device or computer (e.g. shared with family, at public library, etc.)<br>F23. The mobile devices or computers that I usually use are reliable.<br>F24. I can get access to internet connection(s) that are reliable enough for my purposes.<br>F25. I have (for MY USE ALONE) this many mobile devices or computers in my household: (dropdown 0/1/2/more than 2) |

### *4.2.2 Scoring the survey*

Table 12 shows the key to scoring the survey's responses. To enhance concreteness, we introduce three persona names to label the endpoints: Dav, Ash, and Fee. We use these names instead of endpoint labels like "lower SES" to avoid the misleading impression that facet values could somehow predict SES. We assign to persona *Dav* the facet values corresponding to the endpoints shown in foundational literature to be especially prevalent among lower-SES individuals; and assign to *Fee* the opposite endpoints, those especially prevalent among higher-SES individuals. The middle persona, *Ash*, has a mix of facet values. We also remind the reader that these facet values are a matter of individual differences, but that these individual differences do tend to show statistical clustering by SES.

Table 12: SES-Facets Survey key. The more strongly the participant agrees on a question, the closer the facet value is to the endpoint (persona name).

| For Facet Questions: | What does a High score mean: Dav or Fee? |
|---|---|
| F1a, F1b, F1c, F1d | High score → High Self Efficacy (Fee) |
| F2, F3<br>F4 | High score → Risk Averse (Dav)<br>High score → Risk Tolerant (Fee) |
| F7, F10<br>F5, F8 | High score → Low Perceived Control and Authority (Dav)<br>High score → High Perceived Control and Authority (Fee) |
| F14, F13(optional)<br>F11, F12 | High score → High Privacy, Security Concern (Dav)<br>High score → Low Privacy, Security Concern (Fee) |
| F15, F18, F19, F20 | High score → High Communication: Literacy/Education/Culture (Fee) |
| F22<br>F23, F24, F25 | High score → Low Access to Reliable Technology (Dav)<br>High score → High Access to Reliable Technology (Fee) |

The following steps convert the individual results into facet scores:

*Step 1 (Complement/convert when needed):* For all questions except F25, answer scores are numbers from 1 (Completely disagree) to 9 (Completely agree). For some questions, Fee-like answers are closer to 9 (e.g., Question F4), whereas some other questions are the opposite (e.g., Question F3). To make the responses consistently represent Fee-like answers with high numbers, "reverse" the Dav-like questions in Table 12 using the response's 10's complement. (i.e., convert "9" to "1", "8" to "2." and so on). For Question F25, responses need to be transformed: 0 converts to 1, 1 converts to 3, 2 converts to 6 and "More than 2" converts to 9.

*Step 2 (Weighted sum for each facet)*: For each facet (e.g., Attitudes toward Technology Risks), get the weighted sum of the values from Step 1. For example, if we are considering the questions were F2, F3, and F4, the weighted sum would be the response as per Step1 multiplied by the outer weight of that question. $F2 * w_2 + F3 * w_3 + F4 * w_4$ (e.g., $w_2$ is outer weight (0.42) of F2 from Table 6). The outer weights reflect how much each question contributes to the overall assessment of the construct; therefore, the values of the outer weights remain statistically consistent. Each question's outer weights were given in Table 5 to Table 10. This step results in 6 scores per participant, one score for each facet.

If the desired result is a facet score for each participant, then scoring is complete. However, if the desired result is to also categorize each participant's facet values as more Dav-like than Fee-like or vice versa (and/or more Ash-like), Steps 3 and 4 are also needed.

*Step 3 (Calculate facet medians)*: Next, calculate the median value of *each* facet for all participants (e.g., the median Self-Efficacy score, the median Risk score, etc.)

Note that these facet scores are not "absolute." Rather, these are relative to a participant's peer group. For example, a group of college students would be expected to have different levels of computer self-efficacy, risk averse to technology, etc., than a group of retired people. For this purpose, assume that the participants in the study define a peer group. (We believe this to be reasonable, since it has been effective for the GenderMag survey's scoring [Hamid et al. 2024].)

*Step 4 (Tag each participant's facet score)*: To the right of the median (above) is Fee-like, otherwise it is Dav-like. If the participant's facet score is exactly the facet median, then you can choose between Fee or Dav. You can decide in a way that helps balance the sample sizes, or you can add a third tag (Ash-like the persona representing a fraction of target users with SES between Dav's and Fee's).

At the end of Step 4, each participant's facet score is a 6-tuple, one element for each of their facet values. Most participants turn out to have a mix of facet values. For example, a participant might have Technology Self-efficacy, Attitudes toward Technology Risks, and Perceived Control and Authority closer to Dav's; but Privacy/Security, Communication: Literacy/Education/Culture and Access to Reliable Technology closer to Fee's. As Section 5.1 will discuss in detail, these scores calculated from the SES-Facets Survey can be used to analyze not only when a technology is failing to be inclusive, but also what are the reasons, and exactly what types of users (e.g., those with higher concern for privacy/security) are affected by it.

## 4.3 RQ3: Validating Consistency of Our Results with the Foundational Literature

Since the SES facets behind our SES-Facets Survey are based on foundational literature on how SES relates to the individual traits (facet values) in this paper, a validity question for our pair of surveys is whether our participants' responses were consistent with that foundational literature—if not, it would suggest non-generality of the surveys. Hence, the following RQ:

*RQ3-SESAndFacets: Did the distribution of our participants' facet values correspond with their subjective socioeconomic status?*

To make this comparison, we counted lower-SES vs. higher-SES individuals, and also counted Dav-like vs. Fee-like facet values. For the former, we distinguished between lower-SES vs. higher-SES identities using the S-score median value: respondents whose SES-Subjective Survey's S-scores were at or below the median were considered lower-SES and those above the median were considered higher-SES. For the (Dav-like vs. Fee-like), we followed Steps 1-4 of the scoring mechanism in Section 4.2. Thus, for each facet value, we termed the response as Dav-like if it was at or below the median score for that facet value, and if above the median we termed it Fee-like.

Fig. 2 shows the resulting distributions. The x-axis shows each possible combination of the six facet values—from entirely Dav facets (Dav=6, Fee=0) to entirely Fee facets (Dav=0, Fee=6), and every combination between. The y-axis shows the number of Lower-SES vs. Higher-SES participants with that combination of facet values.

As Fig. 2 illustrates, our participants' facet values tended to cluster by SES on both sides of the "middle". Specifically, the number of lower-SES participants and higher-SES participants with equal numbers of Dav-like and Fee-like facet values (under the arrow) is almost identical (63 Lower-SES vs 66 Higher-SES). In contrast, the left side, containing those who had mostly Dav-like facet values, is dominated by lower-SES participants (orange), and the right side is dominated by higher-SES people (blue).

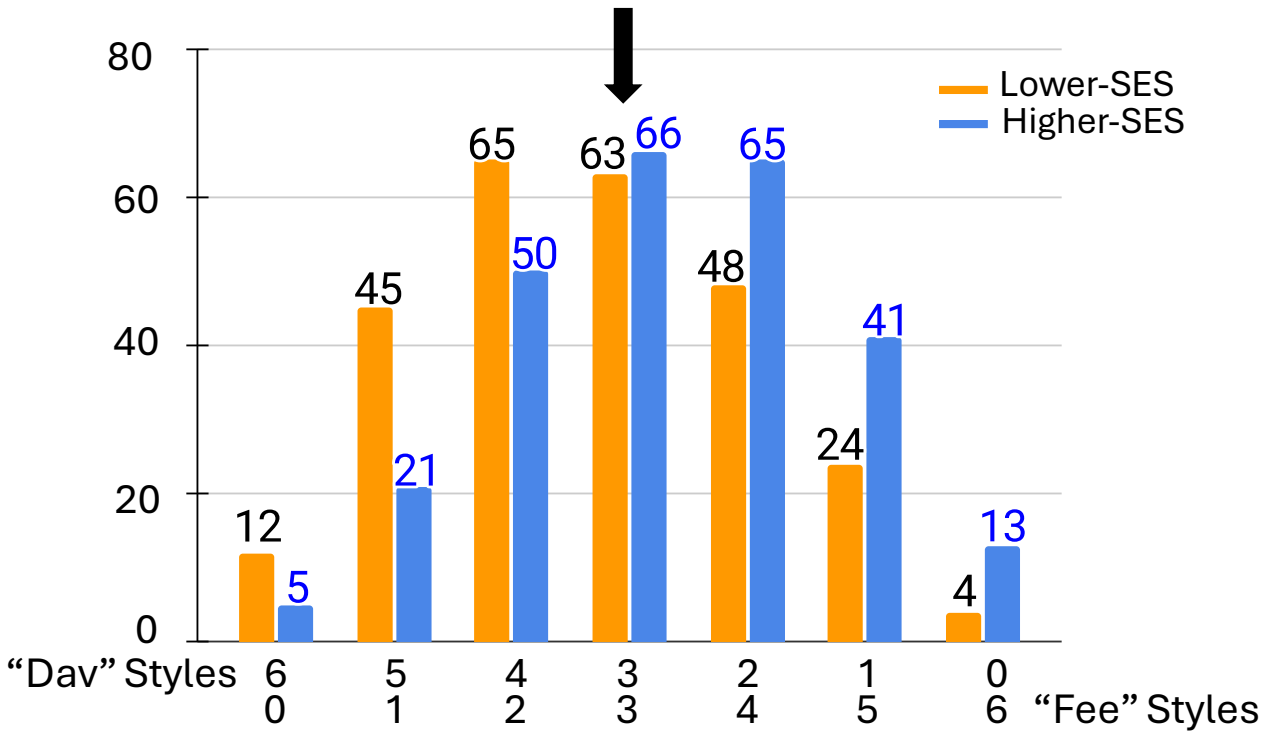

**Fig. 2** Facet values distribution where y-axis is of how many lower-SES and higher-SES participants had the number of facet values shown on the x-axis. Example: the bar at x-axis position "6 0" shows that 12 Lower-SES participants (orange) and 5 Higher-SES participants (blue) had 6 Dav-like facet values and 0 Fee-like facet values. The superimposed arrow points to the participants with an equal number of Dav-like and Fee-like facet values. The Lower-SES participants dominate the "Dav"-like facet values left of the arrow, and the Higher-SES participants dominate the "Fee"-like facet values right of the arrow

Fisher's exact test revealed that the differences were statistically significant. To compute this test, we first excluded the 129 participants who had an equal number of Dav vs. Fee facet values (arrow in Figure 2). We then compared the number of participants with mostly Dav-like facet values vs the number of participants with mostly Fee-like facet values by SES (Table 13). The difference in the counts was significant ($p$<.0001). This significant difference shows that our results are consistent with the foundational literature, which lends credence to the generality of our findings.

Table 13: Fisher's exact test 2x2 contingency table for all possible facet value combinations except 3+3. The number of participants with mostly Dav-like facets vs. those with mostly Fee-like facet values differed significantly by their subjective SES. For Lower-SES participants (orange), the Dav-like facet values dominated, and for Higher-SES participants (blue), the Fee-like facet values dominated. The difference was significant ($p$<.0001).

| | **Mostly Dav facets:** Dav+Fee facet values 6+0, 5+1, & 4+2 | **Mostly Fee facets:** Dav+Fee facet values 2+4, 1+5, 0+6 | Total |
|---|---|---|---|
| Lower-SES | 122 | 76 | 198 |
| Higher-SES | 76 | 119 | 195 |
| Total | 198 | 195 | 393 |
| p < .0001 | | | |

# 5. DISCUSSION

## 5.1 Practical Implications for Software Practitioners

How might software practitioners make practical use of these surveys? We see the SES-Facets Survey as being the most actionable, but the SES-Subjective Survey alone or in combination with the SES-Facets Survey also can bring benefits to software practitioners in several stages of the software lifecycle. The following use-cases illustrate these points.

*Use-Case 1 (Requirements): Requirement gathering (SES-Subjective Survey, SES-Facets Survey):*

Either or both surveys can produce information that informs a software product's requirements. For example, suppose a technology product is targeting a certain geographic area, such as a web-based healthcare information app for city X to match users' healthcare questions to suitable healthcare facilities. Surveying the people in that area can help the software practitioners define the requirements for their web app. The SES-Subjective Survey's results could suggest requirements such as being explicit about monetary costs for the using the service, health insurance prerequisites, and nearby bus stops/transit stations. The SES-Facets Survey's results could further suggest requirements like ensuring that data privacy features are resilient to using shared computers and are compatible with older computer systems. Requirements like these relate to a potential customer's perception of the software usefulness to them, which the TAM/UTAUT body of research has shown to be a critical factor in technology adoption [Williams et al. 2015]. One past example is a Microsoft team's use of the related GenderMag Facet Survey on their customer base, to decide on the pertinence of the GenderMag facet ranges to their product requirements [Hamid et al. 2024].

*Use-Case 2 (Quality Assurance): Recruiting diverse participants for user studies (SES-Subjective Survey, SES-Facets Survey):*

Recruiting a diverse participant pool when performing formative or summative user studies enables evaluating a range of perspectives and experiences of the potential user base. Missing the needs of SES-diverse users can result in products that are biased or widen already existing social inequities. Practitioners could use the SES-Subjective Survey to recruit an SES-diverse user population without needing to ask for their income. The SES-facet survey could help recruit participants who cover wide spectra of SES-facet values. Hilderbrand et al. reported an analogous example of the latter, in which a company used the GenderMag survey to recruit a participant pool balanced with respect to the GenderMag facets [Hilderbrand et al. 2020].

*Use-Case 3 (Quality Assurance): With user studies—to understand the "why" (SES-Facets Survey):*

Supplementing a user study with the SES-Facet Survey can help provide the "*why's"* when participants run into problems with a software feature, because the participants' facet values can suggest the root of that feature's problem. For example, suppose a user study revealed that many users were not using an insurance company's online policy site correctly. Suppose further that the SES-Facet survey then revealed that most of those users heralded from cultures and/or educational backgrounds different than those embedded in the site's communications. This would suggest that the many of the policy site's problems lie in its communications, such as its language usage and its assumptions of what "everyone already knows." Examples demonstrating this kind of use for the GenderMag Facet Survey can be seen in Guizani et al.'s case study with an open source project site's issues recruiting and engaging newcomers [Guizani et al. 2022].

*Use-Case 4 (Debugging): Facets drive fixes (SES-Facets Survey):*

As a follow-on to Use-Case 3 (connecting an inclusivity bug found in a user study with the "Why's" identified from the participants' SES-Facets Survey response), developers could derive fixes from the facet values of participants experiencing that bug. An example using a preliminary version of the SES-Facets Survey can be seen in Agarwal et al.'s field study of a software team debugging a learning management platform product they were building [Agarwal et al. 2023]. In that work, the team addressed issues related to the Communications facet by removing some tech-oriented vocabulary that many participants would have no reason to understand, and adding tooltips to further clarify [Agarwal et al. 2023].

*Use-Case 5 (Maintenance): Comparing versions to evaluate progress (SES-Subjective Survey, SES-Facets Survey):*

Is Version A better than Version B? Multiple user studies on different versions of a product are often performed to answer questions like this. Without the surveys, results tend to be heavily aggregated (e.g., "Version A participants ran into 10% fewer problems than Version B participants.") However, if the SES-Subjective Survey is included in these studies, it can show progress across socioeconomic strata (e.g., "Version A participants ran into 10% fewer problems than Version B participants, and most of the additional gains came because Version A worked better for lower-SES participants.") If the SES-Facets Survey is also included in these studies, the result will be of higher resolution (more detailed) and more actionable (e.g., "all of the low-access-to-reliable-tech participants ran into problems in both versions, but Version A was much more inclusive for the most Privacy-concerned individuals.") The latter's actionability lies in its inclusion of the facet-based Why's, in the same way as with Use-Case 2 and Use-Case 3. Anderson et al. illustrated this kind of usage using the GenderMag Facets survey on 16 AI-powered software products [Anderson et al. 2024].

*Use-Case 6 (Product Reporting): To satisfy reporting requirements (SES-Subjective Survey, SES-Facets Survey):*

The U.S. White House Office of Science and Technology Policy emphasizes the need for technology to avoid "contribut<ing> to unjustified different treatment or impacts disfavoring people based on their <various categories>" [White House 2024], and the European Union has already approved the AI Act, which also includes clauses about biases and inequities [Browne 2024]. As policies like these become federal regulations, many technology companies will need ways to measure the effects of their technology products on different populations, to comply with reporting requirements like these:

> *Reporting. Summary reporting should document ... impacted populations; the assessment of notice clarity... assessment of how explanations are tailored, including to the purpose, the recipient of the explanation, ....*
> [White House 2024]

## 5.2 Threats to Validity

As with any empirical study, our survey study has limitations and threats to validity [Ko et al. 2015; Wohlin et al. 2000]. For this paper, the threats fall into two categories: threats to the validity of the surveys, and threats to the study relating to study design, deployment, data, etc.

The threats to the surveys' validity include the surveys' completeness, their soundness, and their generality. Regarding completeness, survey creators can rarely ask every possible question: instead, they must balance the desire for completeness with participant fatigue. In our case this was particularly true, because many of our intended use-cases suggest that the surveys be used together, so we worked very hard to keep them short while also preserving reliability. In fact, we know that neither survey is a complete set of every question needed to measure (SES-Facets): an individual's socioeconomic facets relevant to their problem-solving on technology, or (SES-Subjective): their subjective SES. We know this about the SES-Facets Survey because the originators of the SES facets, upon which our survey is based, point out that their facet set is a subset of the possible facets [Hu et al. 2021]. We also know it to be true of the SES-Subjective Survey because we pared down the extensive set of questions we harvested from the literature (Appendix B) to the smallest number we could, as we explain in Section 3.1. Thus, this threat is indeed a limitation, but our goal was not completeness, it was validity.

A soundness threat for both surveys, is that the surveys' questions may not adequately measure the constructs we think they measure. We guarded against this threat in several ways. First, the constructs themselves are evidence-based, backed by extensive literature, as explained in Sections 2–4. Second, our indicator reliability analyses showed almost all the questions of both surveys to be reliable measures of the intended construct (the SES facets for the SES-Facets Survey, and subjective SES for the SES-Subjective Survey), with the few exceptions being questions we could not discard for theory reasons. Third, we compared participants' objective SES "ground truth" (as per university data) against our SES-Subjective Survey, and found that the survey results were highly effective at differentiating participants by their objective SES. Finally, we compared the combined surveys' results against predictions made by foundational literature, and again found that our results were consistent with this literature. Given these results, we are very confident of our results' soundness for the populations in our study.

The populations with whom we validated also raise a generality threat that pertains to both our surveys and the study we ran to validate them. Other than their foundational underpinnings, we cannot claim that the surveys are general enough for non-students or non-US individuals. We also cannot claim that the surveys are valid in all pertinent situations. For example, a higher-SES individual who has faced cyberbullying might scrub their personal information from social media (relates to questions F12, F13, F14), a privacy trait more widespread among lower-SES individuals [Marwick et al. 2017]. The same individual might be less concerned about privacy in other types of applications such as for checking prices of online textbooks.

A final threat to our study is that our response rates were between 3% and 4%, which is below the recommended 6% minimum [Smith et al. 2013]. That said, our study's response rates are consistent with some other empirical software engineering publications (e.g., [Gleirscher and Marmsoler 2020, de Mello et al. 2014]). Also, even with our relatively low response rate, we received almost 1000 responses in total (over 400 in one deployment and over 500 in the other). Thus, both surveys' sample sizes are enough for a 5% margin of error for the lower- and higher-SES students at both universities [SurveyMonkey n.d.].

Threats like these can only be addressed by empirically investigating these surveys in additional settings and with additional populations. We have provided the full surveys here, with the hope that interested readers will do so.

# 6. CONCLUSION

*"How well does my software product work for users across a wide socioeconomic spectrum?"* Recent company realization of new market potentials, a rising awareness of ever-widening gaps in society, and several emerging legal requirements all point to a need to answer this question. But how?

We believe that answering it requires *measuring* characteristics of both the product and its intended users. This paper presents two evidence-based surveys for the latter purpose, measuring the users—the SES-Subjective Survey and the SES-Facets Survey—and explains ways to couple these surveys with simultaneous product measurements to answer this question.

- The *SES-Subjective Survey* (RQ1): This survey's primary use, when coupled with product usage data, is to help answer "whether" and "to what extent" questions about a software product's suitability for SES-diverse users. Although other SES-related surveys exist, our SES-Subjective Survey may be particularly well-suited to usage in software engineering, for three reasons. First (1), it focuses on *subjective* SES instead of SES status by some external threshold, because how people view themselves impacts how they behave with (or without) software.

Further (2), it *avoids asking income*, which some participants are unwilling to provide, and some companies are unwilling to ask. Finally (3), it is *general-purpose,* being domain-agnostic and region/country-agnostic.

- The *SES-Facets Survey* (RQ2): As the first survey to measure this set of six individual trait types tied to software usage that statistically vary by SES, it is (1) *unique*. What it adds beyond the SES-Subjective Survey are (2) *high-resolution*, by measuring individuals' six facet values, which brings (3) *actionability*, by revealing specific traits of individuals related to how they work with software. Thus, its primary use, when combined with product usage data, is to answer "why" questions—*why* some particular group of users may not be well-served by some particular software feature.
- The *validations* (RQ1, RQ2, and RQ3): The statistical validations of each survey address both RQ1-SES and RQ2-facets. In addition, the combination of these survey results shows that our data's distribution of facet values among higher- vs. lower-SES individuals is statistically consistent with foundational literature, addressing RQ3-SESAndFacets.

For the in-the-trenches software practitioner, these surveys can serve practical needs across multiple software lifecycle stages. For example, using the SES-Subjective Survey during *Requirements Gathering* could reveal that a product's target population spans a broad spectrum of SES, suggesting that the product needs to support this broad SES spectrum. In contrast, if the survey reveals the opposite, then supporting just one end or the other of the SES spectrum would suffice for that product. During *Software Quality Assurance* activities involving users, either or both surveys could facilitate recruiting a suitably diverse population to cover a SES broad spectrum and/or broad spectra of facet values. Also in *Software Quality Assurance* activities, when coupled with product usage data, the SES-Subjective Survey can answer *who* (by subjective SES) disproportionately runs into problems with the product, and the SES-Facets Survey can add fine-grained insights into *why* (their facet values) these product problems arose for particular groups of users. The *Debugging* and *Maintenance* lifecycle stages are particularly facilitated by the SES-Facets Survey through its explanatory power[1] to answer *why's*, and also enable why-based comparisons of different versions of a single product. Finally, the SES-Subjective Survey enables the kind of *Product Reporting* that can provide evidence of compliance with emerging legislation about AI-powered products.

With these practical applications across the lifecycle in mind, we now return to the question we raised in the Introduction in response to company product visions like Microsoft's:

*How can software practitioners know whether their software products actually do empower a broad spectrum of SES-diverse users?*

We believe our new pair of validated surveys fills a measurement gap key to answering this question. We invite software practitioners across the world to try the surveys out to help answer such questions about their own products.

[1] Following Peirce's definition [Boersema 2003], an explanation has more explanatory power than another if it changes more "surprising facts" (e.g., some group of users running into problems with a software feature...) into "a matter of course" (e.g., ...and those users are highly concerned about their privacy but the feature collects private information).

# Appendix A: SES-Facets Survey as Deployed at University B

*Table A-1: SES-Facets Survey questions deployed at University B.* ***Bolded questions*** *are the remaining questions after the validation in Section 4.2. For all questions except F25 and F26, response scales are from 1 (Completely Disagree) to 9 (Completely Agree). For F25 and F26, dropdown options are added with questions.*

| Technology Self Efficacy |
|---|
| **F1. I am able to use specialized software when...** |
| **a. I have just the built-in help for assistance.** |
| **b. I have seen someone else using it before trying it myself.** |
| **c. someone else has helped me get started.** |
| **d. someone shows me how to do it first.** |

| Attitudes toward Technology Risks |
|---|
| **F2. I avoid "advanced" features or "options" in specialized software.** |
| **F3. I avoid activities when using specialized software that are dangerous or risky.** |
| **F4. Despite the risks, I use features in specialized software that haven't been proven to work.** |

| Perceived Control and Attitude Toward Authority |
|---|
| **F5. I feel I have control over what happens in most situations.** |
| F6. I feel that what happens in my life is often determined by factors beyond my control. |
| **F7. I often have the feeling that I am being treated unfairly.** |
| **F8. I feel I have control over what happens when I work with technology.** |
| F9. When I am working with technology, I feel that what happens is often determined by factors beyond my control. |
| **F10. I often have the feeling that technology ends up treating me unfairly.** |

| Technology Privacy, and Security |
|---|
| **F11. Video cameras in public places that police can view help me feel safe in isolated places/night.** |
| **F12. I do not mind using my full name on social media.** |
| **F13. I delete messages/posts/data from my mobile/social media to guard my personal data.** |
| **F14. I do not want pictures of me on the internet or social media.** |

| Communication: Literacy/ Education/ Culture |
| --- |
| **F15. I usually understand materials referring to USA culture, politics, celebrities, and traditions, (e.g., "it crashed", "ROI", "cutting corners", "FYI", "driving me up the wall" ...).** |
| F16. Some apps/tech I have to use seem to be against the traditions I grew up with. |
| F17. I enjoy reading in general. |
| **F18. When growing up, my school provided me with high quality education.** |
| **F19. When growing up, my school gave me lots of technology to use.** |
| **F20. I can speak English well.** |

| Access to Reliable Technology |
| --- |
| F21. I use mobile devices or computers |
| a. At University |
| b. At Work |
| c. At Home |
| d. Other places (Internet cafe, cyber cafe, etc.) |
| **F22. I use a SHARED mobile device or computer (e.g. shared with family, at public library, etc.)** |
| **F23. The mobile devices or computers that I usually use are reliable** |
| **F24. I can get access to internet connection(s) that are reliable enough for my purposes.** |
| **F25. I have (for MY USE ALONE) this many mobile devices or computers in my household: (dropdown 0/1/2/more than 2)** |
| F26. I SHARE this many mobile devices or computers with others in my household: (dropdown 0/1/2/more than 2) |

# Appendix B: SES-Subjective Survey Questions Literature Sources

*Table B-1: Detailed literature results for papers about SES-related surveys.*

| Row # | Category | SES indicator type | Paper | Concepts asked by the survey / How asked / Respondents / Where deployed |
|---|---|---|---|---|
| 1 | Demographics | Causation | [Omer and Al-Hadithi 2017] | age |
| | | | | Numerical |
| | | | | households in Iraq |
| | | | | Iraq |
| 2 | Demographics | Causation | [Shafiei et al. 2019] | Ethnicity, religion, age, gender, marital status, Household Size, Urban residency, Existence of disabled in the family |
| | | | | Not stated |
| | | | | community of Iran |
| | | | | Iran |
| 3 | Demographics | Causation | [Porter 2016] | 13. sex |
| | | | | female, male |
| | | | | community-based sample of adults (18+) in Dhulikhel, central Nepal *Individuals living in institutionalized settings (e.g. hostels and motels), pregnant women, individuals unable to respond due to a cognitive or physical disability, and those who refused to participate were excluded from the study. |
| | | | | Nepal |
| 4 | Demographics | Causation | [Porter 2016] | 14. age |
| | | | | years: 18-27, 28-37, 38-47, 48-57, 58-67, 68+ |
| | | | | community-based sample of adults (18+) in Dhulikhel, central Nepal *Individuals living in institutionalized settings (e.g. hostels and motels), pregnant women, individuals unable to respond due to a cognitive or physical disability, and those who refused to participate were excluded from the study. |
| | | | | Nepal |
| 5 | Demographics | Causation | [Porter 2016] | 16. Ethnicity |
| | | | | Newar, Brahmin/Chettri/Thakuri/Sanyasi, other |
| | | | | community-based sample of adults (18+) in Dhulikhel, central Nepal *Individuals living in institutionalized settings (e.g. hostels and motels), pregnant women, individuals unable to respond due to a cognitive or physical disability, and those who refused to participate were excluded from the study. |
| | | | | Nepal |
| 6 | Demographics | Causation | [Porter 2016] | 17. religion |
| | | | | Hindu, Buddhist, other |
| | | | | community-based sample of adults (18+) in Dhulikhel, central Nepal *Individuals living in institutionalized settings (e.g. hostels and motels), pregnant women, individuals unable to respond due to a cognitive or physical disability, and those who refused to participate were excluded from the study. |
| | | | | Nepal |
| 7 | Demographics | Causation | [Porter 2016] | 18. marital status |
| | | | | never married, currently married, separated/widowed |
| | | | | community-based sample of adults (18+) in Dhulikhel, central Nepal *Individuals living in institutionalized settings (e.g. hostels and motels), pregnant women, individuals unable to respond due to a cognitive or physical disability, and those who refused to participate were excluded from the study. |
| | | | | Nepal |
| 8 | Demographics | Causation | [Yadegari et al. 2017] | ethnicity |
| | | | | Not stated |
| | | | | pregnant women in their 14 to 42 weeks of pregnancy |
| | | | | city of Rasht, Iran |
| 9 | Demographics: Age | Causation | [Suwannasang 2019] | 2. age |
| | | | | minimum age in house, maximum age in house, median age of respondents |
| | | | | six communities across Bangkok |
| | | | | Nepal |
| 10 | Demographics: Age | Causation | [Babatunde et al. 2007] | 5. age of household head |
| | | | | 18-65, >65 |
| | | | | farming households |
| | | | | Kwara State, North-Central Nigeria |
| 11 | Demographics: Gender | Causation | [Suwannasang 2019] | 3. gender |
| | | | | male, female |
| | | | | six communities across Bangkok |
| | | | | Bangkok, Thailand |
| 12 | Demographics: Gender | Causation | [Babatunde et al. 2007] | 4. gender of household head |
| | | | | male, female |
| | | | | farming households |
| | | | | Kwara State, North-Central Nigeria |
| 13 | Education: Availability Of Information | Causation | [Adamu et al. 2018] | WHOQOL-BREF questionnaire - Availability of information needed for day-to-day life |
| | | | | Not at all 1, Completely 5 |
| | | | | caregivers whose children had been diagnosed with tuberculosis (TB) |
| | | | | Nigeria |
| 14 | Education: Head of Household's Education | Causation | [Roohafza et al. 2021] | "Illiterate, Elementary school, Middle school , Diploma, Bachelor or Associate degree, Master of Sciences or PhD degree, Religious education" |
| | | | | List |
| | | | | Iranian general population |
| | | | | Iran |
| 15 | Education: Head of Household's Education | Causation | [Babatunde et al. 2007] | 3. educational status of household's head |
| | | | | tertiary education, secondary education, primary education, arabic education, no education |
| | | | | farming households |
| | | | | Kwara State, North-Central Nigeria |
| 16 | Education: Head of households' employment status | Causation | [Roohafza et al. 2021] | Governmental, Self-employed, Housewife (for women), Retired, Unemployed, Student |
| | | | | list |
| | | | | Iranian general population |
| | | | | Iran |
| 17 | Education: Household's education | Causation | [Shafiei et al. 2019] | Head of household education, % of literate individuals in family, % of family members who are students, School type (public/private) |
| | | | | Not stated |
| | | | | community of Iran |
| | | | | Iran |
| 18 | Education: Parental Education | Causation | [Raihan et al. 2017] | Education status of the mother: at least one year of formal schooling |
| | | | | Yes/No |
| | | | | Household with children under 5 years old |
| | | | | Bangladesh |
| 19 | Education: Parental Education | Causation | [Pradhan et al. 2018] | Mother's year of schooling from 0 to 16 years are considered. Total year of schooling was divided by 2. If mother have completed the education till elementary they were given score of 1. |
| | | | | 0-8 |
| | | | | household in peri-urban community |
| | | | | Pakistan |
| 20 | Education: Parental Education | Causation | [Ali and Bakar 2019] | Parental education |
| | | | | Not stated |
| | | | | secondary school science students |
| | | | | Pakistan |
| 21 | Education: Parental Education | Causation | [Jordan et al. 2018] | Mother's Education Level:<br>Some graduate education/graduate degree<br>Bachelor's degree<br>Some College<br>High School<br>Less than high school<br>Missing |
| | | | | list |
| | | | | parents of children (age 29 days–18 years at stroke onset) |
| | | | | 9 countries including 4 highest income countries: United States, United Kingdom, Australia, Canada ,3 lower and middle income (LAMI) countries : Philippines, Serbia, and China |
| 22 | Education: Parental Education | Causation | [Colston et al. 2020] | Caregiver's education: whether or not the subject’s caregiver had completed primary education (≥6 completed years of schooling |
| | | | | binary |
| | | | | Households whose children have diarrheal episode or asymptomatic |
| | | | | 11 countries-<br>Bangladesh, Brazil, India, Kenya, Nepal, Pakistan, Peru, South Africa, Tanzania, Zimbabwe, Mali. |
| 23 | Education: Parental Education | Causation | [Carrasco-Escobar et al. 2021] | mothers’ highest educational level |
| | | | | List with options : no education, elementary/primary, secondary, and higher education |
| | | | | Sub-Sahara African community |
| | | | | Sub-Sahara Africa |
| 24 | Education: Parental Education | Causation | [Rosemberg and Alam 2021] | 3. Parent's level of education |
| | | | | 3. Maternal education (in years) |
| | | | | Families of Spanish-speaking Argentinian toddlers (2-3 yrs old) |
| | | | | Buenos Aires, Argentina |
| 25 | Education: Partner's and Personal Education | Causation | [Yadegari et al. 2017] | Couple's education |
| | | | | Not stated |
| | | | | pregnant women in their 14 to 42 weeks of pregnancy |
| | | | | city of Rasht, Iran |
| 26 | Education: Personal Education | Causation | [Omer and Al-Hadithi 2017] | Education levels are divided into 8 levels. The lowest is illiterate which has the value of zero and the highest is doctoral degree (PhD) or equivalent and is given the value of 7. |
| | | | | 8 levels (0-7) |
| | | | | households in Iraq |
| | | | | Iraq |
| 27 | | Causation | [Mustofa et al. 2019] | Educational Levels |
| | Education: Personal Education | | | Not stated<br>graduate students in master program that have scientific publications<br>Indonesia |
| 28 | Education: Personal Education | Causation | [Lee and Yi 2021] | 6. educational level<br>no formal education, elementary school, middle school, or high school or higher<br>women aged 80 and up<br>South Korea |
| 29 | Education: Personal Education | Causation | [Suwannasang 2019] | 4. education<br>A level, bachelor's degree, master's degree, PhD,<br>six communities across Bangkok<br>Bangkok, Thailand |
| 30 | Education: Personal Education | Causation | [Porter 2016] | 15. Years of formal education<br>years: No formal education, Less than high school, High school or more<br>community-based sample of adults (18+) in Dhulikhel, central Nepal *Individuals living in institutionalized settings (e.g. hostels and motels), pregnant women, individuals unable to respond due to a cognitive or physical disability, and those who refused to participate were excluded from the study.<br>Nepal |
| 31 | Education: Personal Educational Attainment | Causation | [Goodwin et al. 2018] | Highest qualification obtained by the participan: no qualifications/GCSE, A-level, degree or above<br>classes/list<br>households from two boroughs in South East London, Lambeth and Southwark<br>UK |
| 32 | Employment | Causation | [Goodwin et al. 2018] | Employment status: Full or part-time employment; student; unemployed; and other. Other employment status included temporary sick, permanently sick or disabled, retired, carer and at home looking after children.<br>classes/list<br>households from two boroughs in South East London, Lambeth and Southwark<br>UK |
| 33 | Employment | Causation | [Porter 2016] | 20. main work status in the past 12 months<br>unemployed, employed, student<br>community-based sample of adults (18+) in Dhulikhel, central Nepal *Individuals living in institutionalized settings (e.g. hostels and motels), pregnant women, individuals unable to respond due to a cognitive or physical disability, and those who refused to participate were excluded from the study. |
| | | | | Nepal |
| 34 | Employment | Causation | [Yadegari et al. 2017] | Employee, Worker, Self-employed |
| | | | | list |
| | | | | pregnant women in their 14 to 42 weeks of pregnancy |
| | | | | city of Rasht, Iran |
| 35 | Employment: Employment status | Causation | [Lee and Yi 2021] | 7. employment status |
| | | | | yes or no |
| | | | | women ages 80 and up |
| | | | | South Korea |
| 36 | Employment: Head of Household's Employment Status | Causation | [Omer and Al-Hadithi 2017] | Job Status: Retired/unemployed/ deceased |
| | | | | Binary |
| | | | | households in Iraq |
| | | | | Iraq |
| 37 | Employment: Household's Employment Status | Causation | [Shafiei et al. 2019] | Head of household job, Second job, % of individuals with employment in family, Unemployment |
| | | | | Not stated |
| | | | | community of Iran |
| | | | | Iran |
| 38 | Food security | Reflection | [Raihan et al. 2017] | Anxiety and uncertainty about the household food supply |
| | | | | Yes/No IF yes then frequency-of-occurrence (Rarely, Sometimes, Often) |
| | | | | Household with children under 5 years old |
| | | | | Bangladesh |
| 39 | Food security | Reflection | [Raihan et al. 2017] | Insufficient Quality (includes variety and preferences of the type of food) |
| | | | | Yes/No IF yes then frequency-of-occurrence (Rarely, Sometimes, Often) |
| | | | | Household with children under 5 years old |
| | | | | Bangladesh |
| 40 | Food security | Reflection | [Raihan et al. 2017] | Insufficient food intake and its physical consequences |
| | | | | Yes/No IF yes then frequency-of-occurrence (Rarely, Sometimes, Often) |
| | | | | Household with children under 5 years old |
| | | | | Bangladesh |
| 41 | Food Security | Reflection | [Yadegari et al. 2017] | Food Security |
| | | | | Yes/No |
| | | | | pregnant women in their 14 to 42 weeks of pregnancy |
| | | | | city of Rasht, Iran |
| 42 | Food Security | Reflection | [Yadegari et al. 2017] | Hunger level |
| | | | | without hunger, moderate hunger, severe hunger |
| | | | | pregnant women in their 14 to 42 weeks of pregnancy |
| | | | | city of Rasht, Iran |
| 43 | Healthcare: Access to Healthcare | Reflection | [Adamu et al. 2018] | WHOQOL-BREF questionnaire - Satisfaction - Access to health services |
| | | | | Very dissatisfied 1, Very satisfied 5 |
| | | | | caregivers whose children had been diagnosed with tuberculosis (TB) |
| | | | | Nigeria |
| 44 | Healthcare: Access to Healthcare | Reflection | [Lee and Yi 2021] | 8. barriers to accessing healthcare |
| | | | | yes, no, or no need for health care |
| | | | | women ages 80 and up |
| | | | | South Korea |
| 45 | Household Size | Causation | [Colston et al. 2020] | Household crowding: whether or not the subject resided in a household with 3 or more residents per bedroom |
| | | | | binary |
| | | | | Households whose children have diarrheal episode or asymptomatic |
| | | | | 12 countries<br>[Bangladesh, Brazil, India, Kenya, Nepal, Pakistan, Peru, South Africa, Tanzania, Zimbabwe, Mali, Mozambique] |
| 46 | Household Size | Causation | [Suwannasang 2019] | 1. average member per house |
| | | | | Number |
| | | | | six communities across Bangkok |
| | | | | Bangkok, Thailand |
| 47 | Household Size | Causation | [Coleman and Fulford 2022] | Number of persons in the household |
| | | | | Not stated |
| | | | | residential addresses within the city of Danville, KY |
| | | | | Danville, Kentucky, USA |
| 48 | Household Size | Causation | [Babatunde et al. 2007] | 2. Household Size |
| | | | | 1-5, 6-10, >10 |
| | | | | farming households |
| | | | | Kwara State, North-Central Nigeria |
| 49 | Household Size | Causation | [Yadegari et al. 2017] | family size |
| | | | | Not stated |
| | | | | pregnant women in their 14 to 42 weeks of pregnancy |
| | | | | city of Rasht, Iran |
| 50 | Household Size | Causation | [Yadegari et al. 2017] | number of family members |
| | | | | Not stated |
| | | | | pregnant women in their 14 to 42 weeks of pregnancy |
| | | | | city of Rasht, Iran |
| 51 | Housing Situation | Reflection | [Patel et al. 2020] | Housing Conditions: Finished floor material, Flush toilet, LPG/electricity for cooking fuel, Improved source of drinking water, More than one room in home |
| | | | | Not stated |
| | | | | households of pregnant women |
| | | | | 6 countries: Guatemala, India, Pakistan, Kenya, Zambia and the Democratic Republic of the Congo |
| 52 | Housing Situation | Reflection | [Colston et al. 2020] | Flooring material: whether or not the subject resided in a household that had a covered ("improved"—rudimentary or finished) as opposed to natural ("unimproved"—earth or sand) floor |
| | | | | binary |
| | | | | Households whose children have diarrheal episode or asymptomatic |
| | | | | 10 countries -<br>Bangladesh, Brazil, India, Kenya, Nepal, Pakistan, Peru, South Africa, Tanzania, Zimbabwe] |
| 53 | Housing Situation | Reflection | [Kasemsant et al. 2019] | 1.1 Status of residence: Owner occupier<br>1.2 Rental Fee per month<br>1.3 Number of rooms in Residence<br>1.4 Number of Bedrooms |
| | | | | 1.1 (Yes or No)<br>1.2 (Baht)<br>1.3 Numerical<br>1.4 Numerical |
| | | | | Thai population |
| | | | | Thailand |
| 54 | Housing Situation | Reflection | [Adamu et al. 2018] | WHOQOL-BREF questionnaire - Ability to get around |
| | | | | Very poor 1, very good 5 |
| | | | | caregivers whose children had been diagnosed with tuberculosis (TB) |
| | | | | Nigeria |
| 55 | Housing Situation: Housing and accommodation status | Reflection | [Shafiei et al. 2019] | Whether the house had a yard, Type of home, Number of rooms, Main cooling devices, Home area, Gas pipe lines, Landline, Toilet, Internet access, Source of water, Electricity, Bathroom, Kitchen, Effluent disposal system, Kind of heating device, House value based on location |
| | | | | Not stated |
| | | | | community of Iran |
| | | | | Iran |
| 56 | Housing Situation: Housing pattern | Reflection | [Lee and Yi 2021] | 2. housing pattern (house, apartment, etc) |
| | | | | traditional house or apartment |
| | | | | women ages 80 and up |
| | | | | South Korea |
| 57 | Housing Situation: Individual's Housing Situation | Reflection | [Adamu et al. 2018] | WHOQOL-BREF questionnaire - Satisfaction - Conditions of living place |
| | | Reflection | [Adamu et al. 2018] | Very dissatisfied 1, Very satisfied 5 |
| | | Reflection | [Adamu et al. 2018] | caregivers whose children had been diagnosed with tuberculosis (TB) |
| | | Reflection | [Adamu et al. 2018] | Nigeria |
| 58 | Housing Situation: Individual's Housing Situation / Personal Space | Reflection | [Roohafza et al. 2021] | # of rooms |
| | | | | List with options : Without room, One, Two, Three and more |
| | | | | Iranian general population |
| | | | | Iran |
| 59 | Housing Situation: Individual's Housing Status | Reflection | [Goodwin et al. 2018] | Own outright/mortgage, private rented, social housing, or rent free; how many times participants had moved in the past 2 years |
| | | | | classes/list |
| | | | | households from two boroughs in South East London, Lambeth and Southwark |
| | | | | UK |
| 60 | Housing Situation: Living arrangement, personal space | Reflection | [Lee and Yi 2021] | 3. living arrangement |
| | | | | living alone, living with a spouse, or living with family or others |
| | | | | women ages 80 and up |
| | | | | South Korea |
| 61 | Housing Situation: Ownership of Household Items | Reflection | [Porter 2016] | 8. Durable goods: Owns a radio, Owns a TV, Owns a mobile phone, Owns a nonmobile phone, Owns a refrigerator, Owns a table, Owns a chair, Owns a sofa, Owns a cupboard, Owns a computer, Owns a clock, Owns a fan, Owns a dhiki or jaato, Owns watch, Owns bike or rickshaw, Owns motorcycle or scooter, Owns car or truck, Has internet, Has a bank account |
| | | Reflection | [Porter 2016] | yes or no |
| | | Reflection | [Porter 2016] | community-based sample of adults (18+) in Dhulikhel, central Nepal *Individuals living in institutionalized settings (e.g. hostels and motels), pregnant women, individuals unable to respond due to a cognitive or physical disability, and those who refused to participate were excluded from the study. |
| | | Reflection | [Porter 2016] | Nepal |
| 62 | Housing Situation: Personal Space | Reflection | [Porter 2016] | 5. Family type |
| | | | | Nuclear, Joint |
| | | | | community-based sample of adults (18+) in Dhulikhel, central Nepal *Individuals living in institutionalized settings (e.g. hostels and motels), pregnant women, individuals unable to respond due to a cognitive or physical disability, and those who refused to participate were excluded from the study. |
| | | | | Nepal |
| 63 | | Reflection | [Porter 2016] | 12. Number of people per sleeping room |
| | | | | Number |
| | Housing Situation: Personal Space | | | community-based sample of adults (18+) in Dhulikhel, central Nepal *Individuals living in institutionalized settings (e.g. hostels and motels), pregnant women, individuals unable to respond due to a cognitive or physical disability, and those who refused to participate were excluded from the study. |
| | | | | Nepal |
| 64 | Housing Situation: Personal Space | Reflection | [Yadegari et al. 2017] | status of home ownership |
| | | | | Not stated |
| | | | | pregnant women in their 14 to 42 weeks of pregnancy |
| | | | | city of Rasht, Iran |
| 65 | Housing Situation: Residential Location | Reflection | [Lee and Yi 2021] | 1. Residential location (urban or rural), |
| | | | | Urban: if the participant lived in a city in the administrative division, Rural: if the respondent lived in a town or township |
| | | | | women ages 80 and up |
| | | | | South Korea |
| 66 | Housing Situation: Residential Location | Reflection | [Jordan et al. 2018] | Urban, Suburban, Rural |
| | | | | list |
| | | | | parents of children (age 29 days–18 years at stroke onset) |
| | | | | 9 countries including 4 highest income countries: United States, United Kingdom, Australia, Canada ,3 lower and middle income (LAMI) countries : Philippines, Serbia, and China |
| 67 | Housing Situation: Residential Location | Reflection | [Le et al. 2015] | Subject's postal code : Open Ended, empty and none Nova Scotia values were excluded |
| | | | | Text |
| | | | | Nova Scotia residents admitted to Queen Elizabeth II burn unit in Halifax, Nova Scotia, from 1995 to 2012 |
| | | | | Nova Scotia, Canada |
| 68 | Housing Situation: Residential Location | Reflection | [Rosemberg and Alam 2021] | 1. Place of residence |
| | | | | Residential neighborhood, Marginalized urban neighborhood |
| | | | | Families of Spanish-speaking Argentinian toddlers (2-3 yrs old) |
| | | | | Buenos Aires, Argentina |
| 69 | Housing Situation: Residential Location | Reflection | [Yadegari et al. 2017] | Location of residence |
| | | | | Not stated |
| | | | | pregnant women in their 14 to 42 weeks of pregnancy |
| | | | | city of Rasht, Iran |
| 70 | Housing Situation: Residential Location | Reflection | [Yadegari et al. 2017] | house area |
| | | | | Not stated |
| | | | | pregnant women in their 14 to 42 weeks of pregnancy |
| | | | | city of Rasht, Iran |
| 71 | Income: Household Income | Causation | [Raihan et al. 2017] | previous month’s income was above 10,000 Tk (Bangladeshi taka) |
| | | | | Yes/No |
| | | | | Household with children under 5 years old |
| | | | | Bangladesh |
| 72 | Income: Household Income | Causation | [Goodwin et al. 2018] | household income in £,<br>Binary variables for current benefit receipt (excluding state pension and child benefit),<br>Debt in the past year (excluding mortgage) |
| | | | | £ & Binary |
| | | | | households from two boroughs in South East London, Lambeth and Southwark |
| | | | | UK |
| 73 | Income: Household Income | Causation | [Pradhan et al. 2018] | Monthly household income in Rupees was obtained in form of range. |
| | | | | Range: Score<br><10,000: 1,<br>10, 000 to 20,000: 2,<br>20,000 to 30,000: 3,<br>>30,000:4<br>score was multiplied by 2 |
| | | | | household in peri-urban community |
| | | | | Pakistan |
| 74 | Income: Household Income | Causation | [Jordan et al. 2018] | <$10,000, $10,000 to $30,000, $30,000 to $50,000, $50,000 to $100,000, and >$100,000, unknown |
| | | | | $ |
| | | | | parents of children (age 29 days–18 years at stroke onset) |
| | | | | 9 countries including 4 highest income countries: United States, United Kingdom, Australia, Canada ,3 lower and middle income (LAMI) countries : Philippines, Serbia, and China |
| 75 | Income: Household Income | Causation | [Lee and Yi 2021] | 4. income level |
| | | | | <1 million KRW,1–2 millionKRW, 2–3 million KRW, or≥3 million KRW |
| | | | | women ages 80 and up |
| | | | | South Korea |
| 76 | Income: Household Income | Causation | [Yadegari et al. 2017] | the amount of rent or mortgage |
| | | | | Not stated |
| | | | | pregnant women in their 14 to 42 weeks of pregnancy |
| | | | | city of Rasht, Iran |
| 77 | Income: Household Income | Causation | [Yadegari et al. 2017] | Income level |
| | | | | < 400 thousands Rial,<br>400 - 800 thousands Rial,<br>> 800 thousands Rial |
| | | | | pregnant women in their 14 to 42 weeks of pregnancy |
| | | | | city of Rasht, Iran |
| 78 | Income: Household Income | Causation | [Yadegari et al. 2017] | total family expenditure |
| | | | | Not stated |
| | | | | pregnant women in their 14 to 42 weeks of pregnancy |
| | | | | city of Rasht, Iran |
| 79 | Income: Household Income | Causation | [Le et al. 2015] | median family household income |
| | | | | 4 levels : Q1: <$49377, Q2: ≥$49377 to <$55885, Q3: ≥$55885, to Q4: >$103564 (Q: quartile) |
| | | | | Nova Scotia residents admitted to Queen Elizabeth II burn unit in Halifax, Nova Scotia, from 1995 to 2012 |
| | | | | Nova Scotia, Canada |
| 80 | Income: Household Income | Causation | [Suwannasang 2019] | 6. income |
| | | | | median household income per month |
| | | | | six communities across Bangkok |
| | | | | Bangkok, Thailand |
| 81 | Income: Household Income | Causation | [Coleman and Fulford 2022] | Household Income |
| | | | | Not stated |
| | | | | residential addresses within the city of Danville, KY |
| | | | | Danville, Kentucky, USA |
| 82 | Income: Parental Income | Causation | [Raihan et al. 2017] | The involvement of mother in income generating activities |
| | | | | Not specified |
| | | | | Household with children under 5 years old |
| | | | | Bangladesh |
| 83 | Income: Personal Income | Causation | [Shafiei et al. 2019] | Monthly income, Monthly saving, Health expenditure, Annual income, Purchasing power, Insurance |
| | | | | Not stated |
| | | | | community of Iran |
| | | | | Iran |
| 84 | Income: Personal Income | Causation | [Adamu et al. 2018] | WHOQOL-BREF questionnaire - Money to meet needs |
| | | | | Not at all 1, Completely 5 |
| | | | | caregivers whose children had been diagnosed with tuberculosis (TB) |
| | | | | Nigeria |
| 85 | Income: Personal Income | Causation | [Mustofa et al. 2019] | Earnings |
| | | | | Not stated |
| | | | | graduate students in master program that have scientific publications |
| | | | | Indonesia |
| 86 | Income: Personal Income | Causation | [Lee and Yi 2021] | 5. current recipient of National Basic Livelihood Security System |
| | | | | NBLSS; yes or no |
| | | | | women ages 80 and up |
| | | | | South Korea |
| 87 | Occupation: Individual's occupation | Causation | [Goodwin et al. 2018] | Occupation: professional & managerial (classes I and II); skilled (class III non-manual and manual); semi-skilled and unskilled (classes IV and V)), |
| | | | | classes/list |
| | | | | households from two boroughs in South East London, Lambeth and Southwark |
| | | | | UK |
| 88 | Occupation: Individual's occupation | Causation | [Omer and Al-Hadithi 2017] | Scores assigned for occupational categories based on the relative importance and prestige of each occupation in Iraq (Unskilled manual occupations, Semi-skilled manual occupations, Skilled manual and non-manual occupations, Associate professional occupations, Skilled professional or senior managerial occupations, and Highly skilled professional occupations) |
| | | | | 1-6 scores |
| | | | | households in Iraq |
| | | | | Iraq |
| 89 | Occupation: Individual's Occupation | Causation | [Suwannasang 2019] | 5. work sector |
| | | | | public, private, own company, |
| | | | | six communities across Bangkok |
| | | | | Bangkok, Thailand |
| 90 | Occupation: Individual's Occupation | Causation | [Porter 2016] | 19. main lifetime occupation |
| | | | | housewife, student, agriculture, sales and service, other |
| | | | | community-based sample of adults (18+) in Dhulikhel, central Nepal *Individuals living in institutionalized settings (e.g. hostels and motels), pregnant women, individuals unable to respond due to a cognitive or physical disability, and those who refused to participate were excluded from the study. |
| | | | | Nepal |
| 91 | Occupation: Parental Occupation | Causation | [Ali and Bakar 2019] | Parental profession |
| | | | | Not stated |
| | | | | secondary school science students |
| | | | | Pakistan |
| 92 | Occupation: Parental Occupation | Causation | [Rosemberg and Alam 2021] | 2. Parent occupation |
| | | | | 2. Professional, Low-qualified jobs, Informal worker, Unemployed |
| | | | | Spanish-speaking Argentinian toddlers (2-3 yrs old) |
| | | | | Buenos Aires, Argentina |
| 93 | Occupation: Power and Authority | Causation | [Mustofa et al. 2019] | Power and Authority |
| | | | | Not stated |
| | | | | graduate students in master program that have scientific publications |
| | | | | Indonesia |
| 94 | Occupation: Prestige | Causation | [Mustofa et al. 2019] | Prestige |
| | | | | Not stated |
| | | | | graduate students in master program that have scientific publications |
| | | | | Indonesia |
| 95 | Standard of Living | Reflection | [Yadegari et al. 2017] | insurance |
| | | | | Not stated |
| | | | | pregnant women in their 14 to 42 weeks of pregnancy |
| | | | | city of Rasht, Iran |
| 96 | Standard of Living | Reflection | [Yadegari et al. 2017] | receiving food aid |
| | | | | Not stated |
| | | | | pregnant women in their 14 to 42 weeks of pregnancy |
| | | | | city of Rasht, Iran |
| 97 | Standard of Living | Reflection | [Yadegari et al. 2017] | whether being supported by social organizations |
| | | | | Not stated |
| | | | | pregnant women in their 14 to 42 weeks of pregnancy |
| | | | | city of Rasht, Iran |
| 98 | Standard of Living | Reflection | [Porter 2016] | 7. Owns livestock: Number of buffalo owned, Number of cows owned, Number of goats owned, Number of sheep owned, Number of chickens owned, Numbers of ducks owned, Number of pigs owned |
| | | | | not stated |
| | | | | community-based sample of adults (18+) in Dhulikhel, central Nepal *Individuals living in institutionalized settings (e.g. hostels and motels), pregnant women, individuals unable to respond due to a cognitive or physical disability, and those who refused to participate were excluded from the study. |
| | | | | Nepal |
| 99 | Standard of Living: Opportunity For Leisure Activities | Reflection | [Adamu et al. 2018] | WHOQOL-BREF questionnaire - The opportunity for leisure activities |
| | | | | Not at all 1, Completely 5 |
| | | | | caregivers whose children had been diagnosed with tuberculosis (TB) |
| | | | | Nigeria |
| 100 | Standard of Living: Opportunity For Leisure Activities | Reflection | [Roohafza et al. 2021] | Fun, pleasure, travel abroad |
| | | | | Yes, No |
| | | | | Iranian general population |
| | | | | Iran |
| 101 | Standard of Living: Opportunity For Leisure Activities | Reflection | [Lee and Yi 2021] | 9. leisure activity |
| | | | | yes or no |
| | | | | women ages 80 and up |
| | | | | South Korea |
| 102 | Standard of Living: Ownership of Communication/ Technology Items | Reflection | [Kasemsant et al. 2019] | 6.1 Number of Computers<br>6.2 Number of Computers connected to internet<br>6.3 Number of Telephones (include PCT trailer)<br>6.4 Number of Mobile phones<br>6.5 Number of Internet users in household |
| | | | | Numerical |
| | | | | Thai population |
| | | | | Thailand |
| 103 | Standard of Living: Ownership of House | Reflection | [Omer and Al-Hadithi 2017] | house ownership |
| | | | | Binary |
| | | | | households in Iraq |
| | | | | Iraq |
| 104 | Standard of Living: Ownership of Household Items | Reflection | [Carrasco-Escobar et al. 2021] | "composite measure based on a Principal Component Analysis (PCA) of a household's cumulative living standard using household's ownership of selected assets, such as televisions and bicycles; materials used for housing construction; and types of water access and sanitation facilities as described elsewhere" |
| | | | | PCA Index |
| | | | | Sub-Sahara African community |
| | | | | Sub-Sahara Africa |
| 105 | Standard of Living: Ownership of Household Items | Reflection | [Yang and Gustafsson 2004] | Car, Central Heat, Computer, Dishwasher, Microwave Oven, Toaster, Stereo, Telephone, Television, Video Recorder /VCR, Video Kee, Clothes Drier, Air Conditioner, Washing Machine, Gas Barbecue, Portable Fire Extinguisher, Bicycle, Cassette Recorder, Radiator, Boat (Sail Or Motor), Own House, Summer House, CD Player, Video Camera, Freezer, Piano/Music Instrument, Encyclopedia, Photographic Equipment, Audio System, Refrigerator, Radio, Satellite Antenna, Laser Record Player, A Second Bathroom, A Second Car, Cordless Phone, Burglar Alarm, Garage, Domestic Help |
| | | | | list of 10 items for each country |
| | | | | 62,413 students total from 23 countries |
| | | | | 23 countries including 2 large countries USA, Canada, another North American country, 17 European countries, 2 Asian countries and New Zealand |
| 106 | Standard of Living: Ownership of Household Items | Reflection | [Raihan et al. 2017] | Have fan, have TV, have electricity, have almirah, have cement and brick walls, have watch clock, have refrigerator |
| | | | | Yes/No |
| | | | | Household with children under 5 years old |
| | | | | Bangladesh |
| 107 | Standard of Living: Ownership of Household Items | Reflection | [Shafiei et al. 2019] | Vacuum cleaner, Washing machine, Dish washing machine, Media player, Hand carpet, Main cooking device, Microwave, Steam-cleaner, Furniture, Camcorder, Radio, dimensional TV (LCD, LED), Color TV, Refrigerator, Side-by-side refrigerator, Freezer, Oven |
| | | | | Not stated |
| | | | | community of Iran |
| | | | | Iran |
| 108 | Standard of Living: Ownership of Household items | Reflection | [Patel et al. 2020] | Household assets: Electricity, Television, Refrigerator, Smart phone, Car,<br>Motorbike, Bicycle |
| | | | | Not stated |
| | | | | household in peri-urban community |
| | | | | 6 countries: Guatemala, India, Pakistan, Kenya, Zambia and the Democratic Republic of the Congo |
| 109 | Standard of Living: Ownership of Household Items | Reflection | [Kasemsant et al. 2019] | 4.1 Number of Wooden or Metal Beds<br>4.2 Number of Gas Stoves<br>4.3 Number of Microwaves<br>4.4 Number of Electronic Vacuum<br>4.5 Number of Refrigerators<br>4.6 Number of Irons<br>4.7 Number of Electronic Pots<br>4.8 Number of Neon Tubes<br>4.9 Number of Compact Fluorescent Lighting |
| | | | | Numerical |
| | | | | Thai population |
| | | | | Thailand |
| 110 | Standard of Living: Ownership of Household Items | Reflection | [Porter 2016] | 1. Type of fuel used for cooking |
| | | | | Liquid petroleum, gas, Wood, Cooks with improved stove |
| | | | | community-based sample of adults (18+) in Dhulikhel, central Nepal *Individuals living in institutionalized settings (e.g. hostels and motels), pregnant women, individuals unable to respond due to a cognitive or physical disability, and those who refused to participate were excluded from the study. |
| | | | | Nepal |
| 111 | Standard of Living: Ownership of Household Items | Reflection | [Porter 2016] | 9. Main material for the floor |
| | | | | Earth or sand, Cement |
| | | | | community-based sample of adults (18+) in Dhulikhel, central Nepal *Individuals living in institutionalized settings (e.g. hostels and motels), pregnant women, individuals unable to respond due to a cognitive or physical disability, and those who refused to participate were excluded from the study. |
| | | | | Nepal |
| 112 | | Reflection | [Porter 2016] | 10. Main material of the roof |
| | Standard of Living: Ownership of Household Items | | | Galvanized sheet, Cement, Ceramic tiles |
| | | | | community-based sample of adults (18+) in Dhulikhel, central Nepal *Individuals living in institutionalized settings (e.g. hostels and motels), pregnant women, individuals unable to respond due to a cognitive or physical disability, and those who refused to participate were excluded from the study. |
| | | | | Nepal |
| 113 | Standard of Living: Ownership of Household Items | Reflection | [Porter 2016] | 11. Main material for external walls |
| | | | | Bricks, Cement, Mud or sand, |
| | | | | community-based sample of adults (18+) in Dhulikhel, central Nepal *Individuals living in institutionalized settings (e.g. hostels and motels), pregnant women, individuals unable to respond due to a cognitive or physical disability, and those who refused to participate were excluded from the study. |
| | | | | Nepal |
| 114 | Standard of Living: Ownership of Household Items | Reflection | [Yadegari et al. 2017] | family economic status (evaluating eight items) |
| | | | | Not stated |
| | | | | pregnant women in their 14 to 42 weeks of pregnancy |
| | | | | city of Rasht, Iran |
| 115 | Standard of Living: Ownership of Household items | Reflection | [Roohafza et al. 2021] | Car Ownership |
| | | | | Yes, No |
| | | | | Iranian general population |
| | | | | Iran |
| 116 | Standard of Living: Ownership of Leisure/Entertainment Items | Reflection | [Kasemsant et al. 2019] | 5.1 Number of Fans/Steam Fans/Ventilators<br>5.2 Number of Radios<br>5.3 Number of TVs<br>5.4 Number of VDO/VCD/DVD players<br>5.5 Number of Washing machines<br>5.6 Number of Air-conditioners<br>5.7 Number of Water heaters in bathroom |
| | | | | Numerical |
| | | | | Thai population |
| | | | | Thailand |
| 117 | Standard of Living: Ownership of Personal Items | Reflection | [Yang and Gustafsson 2004] | Students' own possession: room, (story) books, camera, comic books, magazines/newspaper, place to work, TV, Walkman, video game, video movies, potable stereo, radio, cassette recorder, More than 100 books, encyclopaedia, watch, desk, radio recorder, electronic games, Small radio with headset, pocket disco, bicycle, stereo, mountain bike, rowing boat, tennis racket, slalom skis, telephone, dictionary, typewriter, atlas, Pocket calculator, bed, Books with illustrations |
| | | | | same as #1 |
| | | | | same as #1 |
| | | | | same as #1 |
| 118 | Standard of Living: Ownership of Personal Items | Reflection | [Omer and Al-Hadithi 2017] | car ownership |
| | | | | Binary |
| | | | | households in Iraq |
| | | | | Iraq |
| 119 | Standard of Living: Ownership of Personal Items | Reflection | [Shafiei et al. 2019] | Ownership of car, Motorcycle, Bicycle, Mobile, Smart phone, Personal computer/laptop |
| | | | | Not stated |
| | | | | community of Iran |
| | | | | Iran |
| 120 | Standard of Living: Ownership of Personal Items | Reflection | [Pradhan et al. 2018] | A total of eight priority assets* out of 25 were selected depicting SES standing in the local context. For each asset, households were assigned 1 if they have the asset and 0 if they do not have the asset. These scores were then summed.<br>1. TV<br>2. Refrigerator<br>3. Air conditioner<br>4. Computer<br>5. Net connection<br>6. Possession of land<br>7. Any pet animal (goat or cow)<br>8. Car//truck |
| | | | | 0-8 |
| | | | | household in peri-urban community |
| | | | | Pakistan |
| 121 | Standard of Living: Ownership of Personal Items | Reflection | [Roohafza et al. 2021] | Using notebook, laptop, or tablet in the house |
| | | | | Yes, No |
| | | | | Iranian general population |
| | | | | Iran |
| 122 | Standard of Living: Ownership of Personal Items | Reflection | [Kasemsant et al. 2019] | 3.1 Number of Motorcycles<br>3.2 Number of Personal cars<br>3.3 Number of Pickups /Vans/Mini-vans |
| | | | | Numerical |
| | | | | Thai population |
| | | | | Thailand |
| 123 | Standard of Living: Ownership of Real Estate | Reflection | [Porter 2016] | 6. Owns agricultural land: Amount of land owned (sq. ft) |
| | | | | not stated |
| | | | | community-based sample of adults (18+) in Dhulikhel, central Nepal *Individuals living in institutionalized settings (e.g. hostels and motels), pregnant women, individuals unable to respond due to a cognitive or physical disability, and those who refused to participate were excluded from the study. |
| | | | | Nepal |
| 124 | | Reflection | | 6. farm size |
| | Standard of Living: Ownership of Real Estate | | [Babatunde et al. 2007] | 0.1-2.5, >2.5 |
| | | | | farming households |
| | | | | Kwara State, North-Central Nigeria |
| 125 | Standard of Living: Standard of Living | Reflection | [Kasemsant et al. 2019] | 2.1 Type of fuel that use for<br>2.2 Bottled or Tap Water<br>2.3 Flushing Lavatory |
| | | | | 2.1 Gas or Coal<br>2.2 Bottled or Tap Water<br>2.3 Numerical |
| | | | | Thai population |
| | | | | Thailand |
| 126 | Transportation | Reflection | [Adamu et al. 2018] | WHOQOL-BREF questionnaire - Satisfaction - transport |
| | | | | Very dissatisfied 1, Very satisfied 5 |
| | | | | caregivers whose children had been diagnosed with tuberculosis (TB) |
| | | | | Nigeria |
| 127 | Water/Sanitation | Reflection | [Pradhan et al. 2018] | Improved Water Using WHO definition of improved water, households with improved water include (1. Municipal network 2. Private well 3. Bottled Water (bottled water is considered improved only when the household use another improved source for cooking and personal hygiene) 4. Boring)<br><br>Improved Sanitation Using WHO definition of improved sanitation below categorization was considered as improved sanitation. (1. Flush to piped sewer system 2. Ventilated improved pit latrine 3. Flush to septic tank 4. Pit latrine with slab) |
| | | | | 0-8 |
| | | | | household in peri-urban community |
| | | | | Pakistan |
| 128 | Water/Sanitation: Sanitation | Reflection | [Colston et al. 2020] | Sanitation: whether or not the subject resided in a household with access to an improved, non-shared sanitation facility ("improved" meaning designed to hygienically separate excreta from human contact) |
| | | | | binary |
| | | | | Households whose children have diarrheal episode or asymptomatic |
| | | | | 9 countries -<br>Bangladesh, Brazil, India, Kenya, Nepal, Pakistan, Peru, South Africa, Tanzania |
| 129 | Water/Sanitation: Standard of | Reflection | [Porter 2016] | 3. Main source of drinking water |
| | | | | Drinking water piped into dwelling |
| | Living indicators - Drinking water | | | community-based sample of adults (18+) in Dhulikhel, central Nepal *Individuals living in institutionalized settings (e.g. hostels and motels), pregnant women, individuals unable to respond due to a cognitive or physical disability, and those who refused to participate were excluded from the study. |
| | | | | Nepal |
| 130 | Water/Sanitation: Standard of Living indicators - Drinking water | Reflection | [Porter 2016] | 4. Treatment methods of drinking water |
| | | | | Water treatment method: Water filter, Water treatment method: Boil |
| | | | | community-based sample of adults (18+) in Dhulikhel, central Nepal *Individuals living in institutionalized settings (e.g. hostels and motels), pregnant women, individuals unable to respond due to a cognitive or physical disability, and those who refused to participate were excluded from the study. |
| | | | | Nepal |
| 131 | Water/Sanitation: Standard of Living indicators - type of toilet | Reflection | [Porter 2016] | 2. Type of toilet |
| | | | | Flush to septic tank, Flush to piped sewerage system, Flush to pit latrine, Uses a shared toilet |
| | | | | community-based sample of adults (18+) in Dhulikhel, central Nepal *Individuals living in institutionalized settings (e.g. hostels and motels), pregnant women, individuals unable to respond due to a cognitive or physical disability, and those who refused to participate were excluded from the study. |
| | | | | Nepal |
| 132 | Water/Sanitation: Water Sources | Reflection | [Colston et al. 2020] | Drinking water sources: whether or not the subject resided in a household with access to an improved drinking water source (with potential to deliver safe water by nature of its design and construction such as piped water or protected tubewells, boreholes, dug wells, or springs) |
| | | | | binary |
| | | | | Households whose children have diarrheal episode or asymptomatic |
| | | | | 8 countries with low resource settings - Bangladesh, Brazil, India, Nepal, Pakistan, Peru, South Africa, Tanzania |
| 133 | Well-being: Ability to Concentrate | Reflection | [Adamu et al. 2018] | WHOQOL-BREF questionnaire - Ability to concentrate |
| | | | | Not at all 1, Extremely 5 |
| | | | | caregivers whose children had been diagnosed with tuberculosis (TB) |
| | | | | Nigeria |
| 134 | Well-being: Ability To | Reflection | [Adamu et al. 2018] | WHOQOL-BREF questionnaire - Satisfaction - Ability to perform daily living activities |
| | | | | Very dissatisfied 1, Very satisfied 5 |
| | Perform Daily Activities | | | caregivers whose children had been diagnosed with tuberculosis (TB) |
| | | | | Nigeria |
| 135 | Well-being: Accepting Bodily Appearance | Reflection | [Adamu et al. 2018] | WHOQOL-BREF questionnaire - accepting bodily appearance |
| | | | | Not at all 1, Completely 5 |
| | | | | caregivers whose children had been diagnosed with tuberculosis (TB) |
| | | | | Nigeria |
| 136 | Well-being: Capacity for Work | Reflection | [Adamu et al. 2018] | WHOQOL-BREF questionnaire - Satisfaction - capacity for work |
| | | | | Very dissatisfied 1, Very satisfied 5 |
| | | | | caregivers whose children had been diagnosed with tuberculosis (TB) |
| | | | | Nigeria |
| 137 | Well-being: Energy for Everyday Life | Reflection | [Adamu et al. 2018] | WHOQOL-BREF questionnaire - energy for everyday life |
| | | | | Not at all 1, Completely 5 |
| | | | | caregivers whose children had been diagnosed with tuberculosis (TB) |
| | | | | Nigeria |
| 138 | Well-being: Enjoying Life | Reflection | [Adamu et al. 2018] | WHOQOL-BREF questionnaire - enjoying life |
| | | | | Not at all 1, An extreme amount 5 |
| | | | | caregivers whose children had been diagnosed with tuberculosis (TB) |
| | | | | Nigeria |
| 139 | Well-being: Feeling Safe | Reflection | [Adamu et al. 2018] | WHOQOL-BREF questionnaire - Feeling safe |
| | | | | Not at all 1, Extremely 5 |
| | | | | caregivers whose children had been diagnosed with tuberculosis (TB) |
| | | | | Nigeria |
| 140 | Well-being: Health of Physical Environment | Reflection | [Adamu et al. 2018] | WHOQOL-BREF questionnaire - Health of physical environment |
| | | | | Not at all 1, Extremely 5 |
| | | | | caregivers whose children had been diagnosed with tuberculosis (TB) |
| | | | | Nigeria |
| 141 | Well-being: Meaning of life | Reflection | [Adamu et al. 2018] | WHOQOL-BREF questionnaire - Meaing of life |
| | | | | Not at all 1, An extreme amount 5 |
| | | | | caregivers whose children had been diagnosed with tuberculosis (TB) |
| | | | | Nigeria |
| 142 | | Reflection | [Adamu et al. 2018] | WHOQOL-BREF questionnaire - Need for medical treatment |
| | Well-being: Need or Medical Treatment | | | Not at all 1, An extreme amount 5<br>caregivers whose children had been diagnosed with tuberculosis (TB)<br>Nigeria |
| 143 | Well-being: Negative Feelings | Reflection | [Adamu et al. 2018] | WHOQOL-BREF questionnaire - negative feelings<br>Never 1, Always 5<br>caregivers whose children had been diagnosed with tuberculosis (TB)<br>Nigeria |
| 144 | Well-being: Personal Health Satisfaction | Reflection | [Adamu et al. 2018] | WHOQOL-BREF questionnaire - Health satisfaction<br>Very dissatisfied 1, Very satisfied 5<br>caregivers whose children had been diagnosed with tuberculosis (TB)<br>Nigeria |
| 145 | Well-being: Personal Relationships Satisfaction | Reflection | [Adamu et al. 2018] | WHOQOL-BREF questionnaire - Satisfaction - Personal relationships<br>Very dissatisfied 1, Very satisfied 5<br>caregivers whose children had been diagnosed with tuberculosis (TB)<br>Nigeria |
| 146 | Well-being: Physical Pain | Reflection | [Adamu et al. 2018] | WHOQOL-BREF questionnaire - Physical pain<br>Not at all 1, An extreme amount 5<br>caregivers whose children had been diagnosed with tuberculosis (TB)<br>Nigeria |
| 147 | Well-being: Quality Of Life | Reflection | [Adamu et al. 2018] | WHOQOL-BREF questionnaire - quality of life<br>Very poor 1, very good 5<br>caregivers whose children had been diagnosed with tuberculosis (TB)<br>Nigeria |
| 148 | Well-being: Self Satisfaction | Reflection | [Adamu et al. 2018] | WHOQOL-BREF questionnaire - Satisfaction - Self<br>Very dissatisfied 1, Very satisfied 5<br>caregivers whose children had been diagnosed with tuberculosis (TB)<br>Nigeria |
| 149 | Well-being: Sex Life Satisfaction | Reflection | [Adamu et al. 2018] | WHOQOL-BREF questionnaire - Satisfaction - sex life<br>Very dissatisfied 1, Very satisfied 5<br>caregivers whose children had been diagnosed with tuberculosis (TB)<br>Nigeria |
| 150 | Well-being: Sleep Satisfaction | Reflection | [Adamu et al. 2018] | WHOQOL-BREF questionnaire - Satisfaction - Sleep<br>Very dissatisfied 1, Very satisfied 5<br>caregivers whose children had been diagnosed with tuberculosis (TB) |
| | | | | Nigeria |
| 151 | Well-being: Support From Friends | Reflection | [Adamu et al. 2018] | WHOQOL-BREF questionnaire - Satisfaction - Support from friends |
| | | | | Very dissatisfied 1, Very satisfied 5 |
| | | | | caregivers whose children had been diagnosed with tuberculosis (TB) |
| | | | | Nigeria |